\documentclass[usenatbib]{mn2e}
\usepackage{txfonts}
\usepackage{graphicx}
\usepackage{natbib}
\usepackage{times}
\usepackage{epsfig,rotating}
\usepackage{rotating}

\def \al {\rm et al.~}
\def \13CO {$^{13}$CO}
\def \C18O {C$^{18}$O}

\def \arcsec {$^{\prime\prime}$~}
\def \arcmin {$^\prime$}

\def \3P1 {$^3$P$_1$ -- $^3$P$_0$}

\def \H2 {H$_{2}$}
\def \nh2 {n$_{H_2}$\,}

\def \degree {$\ensuremath{^\circ}\,$}

\begin{document}
\title{A deep ATCA 20cm radio survey of the AKARI Deep Field South near the South Ecliptic Pole}
\author[Glenn J. White et al.]
{Glenn J. White$^{1,2}$, Bunyo Hatsukade$^3$, Chris Pearson$^{2,1,6}$,Toshinobu Takagi$^4$, 
\newauthor Chris Sedgwick$^1$, Shuji Matsuura$^4$, Hideo Matsuhara$^4$, Stephen Serjeant$^1$,
\newauthor Takao Nakagawa$^4$, Hyung Mok Lee$^5$, Shinki Oyabu$^{13}$, Woong-Seob Jeong$^{10}$, 
\newauthor Mai Shirahata$^{4,12}$, Kotaro Kohno$^{7,8}$, Issei Yamamura$^4$, Hitoshi Hanami$^9$, 
\newauthor Tomotsugu Goto$^{11}$, Sin'itirou Makiuti$^4$, David L. Clements$^{14}$, Malek, K.$^{15}$,
\newauthor Sophia A. Khan$^{16}$ \\\\
$^1$ Department of Physics and Astronomy, The Open University, Walton Hall, Milton Keynes, MK7 6AA, UK\\
$^2$ RAL Space, STFC Rutherford Appleton Laboratory, Chilton, Didcot, Oxfordshire, OX11 0QX, UK\\
$^3$ Department of Astronomy, Kyoto University, Kyoto 606-8502, Japan\\
$^4$ Institute of Space and Astronautical Science, JAXA, 3-1-1 Yoshinodai, Chuo, Sagamihara, Kanagawa 252-5210, Japan,\\
$^5$ Astronomy Program, Department of Physics and Astronomy, Seoul National, University, Seoul 151-747, Korea\\
$^6$ Institute for Space Imaging Science, University of Lethbridge,, Lethbridge, Alberta T1K 3M4, Canada\\
$^7$ Institute of Astronomy, the University of Tokyo, 2-21-1 Osawa, Mitaka, Tokyo 181-0015, Japan\\
$^8$ Research Center for the Early Universe, University of Tokyo, 7-3-1 Hongo, Bunkyo, Tokyo 113-0033, Japan\\
$^9$Physics Section, Faculty of Humanities and Social Sciences, Iwate University, Morioka 020-8550, Japan\\
$^{10}$KASI, 61-1, Whaam-dong, Yuseong-gu, Deajeon, 305-348, South Korea\\
$^{11}$Institute for Astronomy, University of Hawaii, 2680 Woodlawn Drive, Honolulu, HI, 96822, USA\\
$^{12}$Subaru Telescope, National Astronomical Observatory of Japan, 650 North A'ohoku Place, Hilo, HI, 96720, U.S.A\\
$^{13}$Graduate School of Science, Nagoya University, Furo-cho, Chikusa-ku, Nagoya, Aichi 464-8602, Japan\\
$^{14}$Imperial College, London, Blackett Lab, Prince Consort Road, London SW7 2AZ, UK\\
$^{15}$Center for Theoretical Physics of the Polish Academy of Sciences, Al. Lotnikow 32/46, 02-668 Warsaw, Poland\\
$^{16}$Shanghai Key Lab for Astrophysics, Shanghai Normal University, Shanghai 200234, China}
\maketitle

\begin{abstract}\\
The results of a deep radio survey at 20 cm wavelength are reported for a region containing the AKARI Deep Field South (ADF-S) near the South Ecliptic Pole (SEP), using the Australia Telescope Compact Array telescope, ATCA. The survey (hereafter referred to as the ATCA-ADFS survey) has 1$\sigma$ detection limits ranging from 18.7--50 $\mu$Jy beam$^{-1}$ over an area of  $\sim$ 1.1 degree$^2$, and $\sim$ 2.5 degree$^2$ to lower sensitivity. The observations, data reduction and source count analysis are presented, along with a description of the overall scientific objectives, and a catalogue containing 530 radio sources detected with a resolution of 6.2\arcsec $\times$ 4.9$^{\prime\prime}$. The derived differential source counts  show a pronounced excess of sources fainter than $\sim$ 1 mJy, consistent with an emerging population of star forming galaxies. Cross-correlating the radio with AKARI sources and archival data we find 95 cross matches, with most galaxies having optical R-magnitudes in the range 18-24 magnitudes, similar to that found in other optical deep field identifications, and 52 components lying within 1\arcsec ~of a radio position in at least one further catalogue (either IR or optical). We have reported redshifts for a sub-sample of our catalogue finding that they vary between galaxies in the local universe to those having redshifts of up to 0.825. Associating the radio sources with the Spitzer catalogue at 24 $\mu$m, we find 173 matches within one Spitzer pixel, of which a small sample of the identifications are clearly radio loud compared to the bulk of the galaxies. The radio luminosity plot and a colour-colour analysis suggest that the majority of the radio sources are in fact luminous star forming galaxies, rather than radio-loud AGN. There are additionally five cross matches between ASTE or BLAST submillimetre galaxies and radio sources from this survey, two of which are also detected at 90 $\mu$m, and 41 cross-matches with submillimetre sources detected in the ${\it Herschel}$ HerMES survey Public Data release.
\end{abstract}

radio continuum: galaxies; surveys

\section{Introduction}

A fundamental challenge  in contemporary astrophysics is to understand how the galaxies have evolved to their current form. To address this issue, wide area surveys are required to accumulate large statistical samples of galaxies. To study this question, the Japanese AKARI infrared satellite (Murakami \al 2007)  carried out two deep infrared legacy surveys close to the North and South Ecliptic Poles (Matsuhara \al 2006, Matsuura \al 2011), which are notable because their sight-lines to the distant Universe have the advantages of low extinction and correspondingly small Hydrogen column densities. To support the two AKARI Deep Fields,  sensitive radio surveys have been made of both ecliptic pole regions to study and compare the global properties of the extragalactic source populations (White \al 2009, White \al 2010a [hereafter 'Paper 1']). In the present paper the results are reported of a sensitive radio survey at 1.4 GHz using the Australia Telescope Compact Array (ATCA) of a region that includes both the ADF-S field (Matsuhara \al 2006, Wada \al 2008, Shirahata \al 2009, White \al 2009, Matsuura \al 2009, 2011), as well as a more extended region around it. The ADF-S is the focus of a major multi-wavelength observing campaign conducted across the entire spectral region. The combination of these far-infrared data and the depth of the radio observations will allow unique studies of a wide range of topics including the redshift evolution of the luminosity function of radio sources, the clustering environment of radio galaxies, the nature of obscured radio-loud Active Galactic Nuclei (AGN), and the radio/far-infrared correlation for distant galaxies. 

\section{Multi-wavelength observations}

The AKARI  ADF-S field is a region located close to the South Ecliptic Pole (Matsuura \al 2009, 2011) with a very low cirrus level $\le$ 0.5 MJy sr$^{-1}$ (Schlegel \al 1998, Bracco \al 2011), and correspondingly low Hydrogen column density $\sim$ 5$\times$10$^{19}$ cm$^{-2}$. This field is similar to the well known Lockman Hole and Chandra Deep-Field South regions, and has half of the cirrus emission of the well studied COSMOS field at 24$\mu$m. The ADF-S field is therefore one of the best 'cosmological windows' through which to study the distant Universe (Malek \al 2009, Matsuura \al 2011, Hajian \al 2012), and is now of high priority for astronomers to build ancillary data sets that can be compared with the AKARI data, and to prepare lead on to the next set of deep cosmological surveys, such as those that will be provided by ${\it Herschel}$ (Pilbratt \al 2010) and SPICA (Eales \al 2009, Swinyard \al 2009).

The AKARI ADF-S survey was primarily made in the far-infrared at wavelengths of 65, 90, 140, 160 $\mu$m over a 12 deg$^2$ area with the AKARI Far-Infrared Surveyor (FIS) instrument (Kawada  \al 2007), with shallower mid-infrared coverage at 9, 18$\mu$m using the AKARI Infrared Camera (IRC) instrument (Onaka \al 2007). In addition to the wide survey, deeper mid-infrared pointed observations, using the IRC, covering $\sim$ 0.8 deg$^2$ and reaching 5$\sigma$ sensitivities of 16, 16, 74, 132, 280 and 580 $\mu$Jy at 3.2, 4.6, 7, 11, 15, 24 $\mu$m were also carried out.  At other wavelengths, the region has recently been mapped by Spitzer's Multi-band imaging photometer (MIPS) at 24 and 70 $\mu$m (Scott \al 2010, Clements \al 2011); by the Balloon-borne Large Aperture Submillimeter Telescope (BLAST) at 250, 350 and 500 $\mu$m (Valiante \al 2010), the latter revealing $\sim$200 sub-millimetre galaxies over an 8.5 deg$^2$ field; and in the ground-based submillimetre band by Hatsukade \al (2011) revealing 198 potential sub-millimetre galaxies in an $\sim$0.25 square degree area. The ancillary data sets summarised in Table \ref{ancillary} will be used in calibration of the radio positional reference frame, and for cross-identifications later in this paper. The AKARI sensitivity limits correspond approximately to being able to detect starburst galaxies and AGN with a luminosity of 10$^{12}$ ${\it L}_{\odot}$ at ${\it z}$ = 0.5, or ultraluminous infrared galaxies (ULIRGS) with luminosities 10$^{12-13}$ ${\it L}_{\odot}$ at ${\it z}$ = 1--2 respectively. Note that the ADF-S has also been observed by the ${\it Herschel}$ Space Observatory (HSO) (Pilbratt \al 2010) as part of the Herschel Multi-tiered Extragalactic Survey (HerMES) guaranteed time key program (Oliver \al 2010).

Optical, radio, X-ray and infrared surveys provide essential support to the interpretation of deep extragalactic radio surveys. The ADF-S has been the focus of recent multi-wavelength survey coverage by our team, with optical imaging with the CTIO 4m telescope (MOSIAC-II detector) to an R-band sensitivity of 25 magnitudes, and at near-IR wavelengths to K $\sim$18.5 magnitudes with the IRSF/SIRIUS instrument already completed. To support the ADF-S and ATCA surveys, we have separately obtained wide field imaging in the optical and near-IR at ESO (using WFI and SOFI), at the AAT (using WFI and IRIS2), for fields of 0.5 - 1 square degree, and spectroscopic observations using AAOmega on the AAT (Sedgwick \al 2009, 2011).

\begin{table*}
\vspace{0pt}
\caption{Summary of ancillary observations available for the ATCA-ADFS deep field}
\begin{scriptsize}
\fontsize{8}{10}\selectfont
\begin{tabular}{l l l l l}
\hline
\multicolumn{1}{l}{Wavelength} & \multicolumn{1}{c}{Telescope} & \multicolumn{1}{c}{Area} & \multicolumn{1}{c}{Beam size} & \multicolumn{1}{c}{Depth}\\ 
\multicolumn{1}{l}{(1)} & \multicolumn{1}{c}{(2)} & \multicolumn{1}{c}{(3)} & \multicolumn{1}{c}{(4)} &  \multicolumn{1}{c}{(5)} \\
\hline
FUV, NUV & GALEX & Central 1 degree$^{2}$ & 6\arcsec & 25.5-26.5 mag (AB)\\
R &  CTIO MOSAIC-II & 7.2 degree$^2$ & 1\arcsec & R 25 mag\\ 
U, B, V, I & CTIO / MOSAIC & Central 1 degree$^2$ & 1\arcsec & U 25, B 26, V 26, I 25 mag\\ 
3-24~$\mu$m (6 bands) & AKARI / IRC & Central 0.8 degree$^2$ & 4.2 - 5.5\arcsec & 10 $\mu$Jy @3.5$\mu$m\\ 
& & & &300uJy @15�m\\ 
24~$\mu$m $\&$ 70~$\mu$m & ${\it Spitzer}$-MIPS & 11 degree$^2$ & 6\arcsec-18\arcsec & 200 $\mu$Jy @24$\mu$m\\ 
& & & &20~mJy @70$\mu$m\\ 
65~$\mu$m, 90~$\mu$m, 140~$\mu$m, 160~$\mu$m & AKARI / FIS & 12 degree$^2$ & 37\arcsec~--~ 50\arcsec & 30mJy (3$\sigma$) @90$\mu$m\\ 
110-500~$\mu$m & ${\it Herschel}$ (HerMES GT) & 7 degree$^2$ & 8\arcsec~ -- 36\arcsec & 30 mJy (5$\sigma$)\\
250~$\mu$m, 350~$\mu$m, 500~$\mu$m & BLAST & 9 degree$^2$ & 36\arcsec~-- 60\arcsec & 45mJy (3$\sigma$) all bands\\ 
870~$\mu$m & APEX / LABOCA & Central 20\arcmin$\times$20\arcmin & 19\arcsec & 6 mJy (3$\sigma$)\\ 
1.1 mm & ASTE / AzTEC & Central 0.25 degree$^2$ & 30\arcsec & 1.2-2.4 mJy (3$\sigma$)\\ 
20 cm & ATCA & Central 1 degree$^2$ & 10\arcsec & 17 $\mu$Jy\\ 
Spectroscopy & AAT  AAOmega & Central 3.14 degree$^2$ &-& R$\sim$21 mag\\ 
Spectroscopy & IMACS  Magellan & Central 0.2 degree$^2$ &-& I$\sim$22 mag\\ 
\hline
 \end{tabular}
\end{scriptsize}
\label{ancillary}
\end{table*}

\section{Radio Observations}
\label{section:observations}

\subsection{ATCA observations}
The radio observations were collected over a 13 day period in July 2007 using the ATCA operated at 1.344 and 1.432 GHz. The total integration time for the 2007 observations was 120 h, spread between 26 overlapping pointing positions to maximise the uv coverage and to mitigate the effects of sidelobes from nearby radio-bright sources. Two of the pointing positions were observed on each night, by taking one five minute integration at each of the two target fields, followed by a two minute integration on the nearby secondary calibrator 0407-658. This cycle was repeated for the different  pointing positions, which were observed over $\sim$10 hour tracks each night, giving similar uv-coverage for each target field. The amplitude scaling was bootstrapped from the primary calibrator PKS 1934-638, which was observed for 10 min at the start of each observing night, and which was assumed to have a flux density of 15.012 Jy at 1.344 GHz and 14.838 Jy at 1.432 GHz respectively. The 2007 data were augmented with a further deep observation made in December 2008  over 5 nights toward a single pointing position at the ADF-S, which lay just off centre of the larger ATCA-ADFS field reported here. This added a further 50 hours of integration time. The data were processed in exactly the same way as that from the 2007 observing sessions.

\subsection{Calibration}

In the following sub-sections the calibration and data reduction methodology are presented. Since much of this is in common with our recent North Ecliptic Pole (NEP) radio survey with the Westerbork Synthesis Radio Telescope (WSRT) telescope discussed in Paper 1, we will not repeat the detailed discussion of this earlier paper, but instead just focus on those parts of the calibration methodology that differed from Paper 1.

The data were calibrated using the ATNF data reduction package ${\it MIRIAD}$ (Sault \al 1995) using standard procedures. The raw data come in RPFITS format, and were converted into the native ${\it MIRIAD}$ format using ATLOD. ATLOD discards every other frequency channel (since they are not independent from one another, hence no information is lost), and additionally flagged out one channel in the higher frequency sideband which contained a multiple of 128 MHz, and thus was affected by self-interference at the ATCA. Channels at either end of the sidebands where the sensitivity dropped significantly were also not used. The resulting data set contained two sidebands, with 13  and 12 channels respectively, each 8 MHz wide, which resulted in a total bandwidth of 200 MHz. The lower frequency sideband was mostly free of RFI and required little editing apart from flagging of bad data. However, the higher frequency sideband suffered from occasional local RF interference, and the affected data were flagged out using the ATNF automated noise flagger PIEFLAG (Middelberg 2006), which eliminated virtually all of the RFI-affected data which would have been flagged in a visual inspection. A visual inspection of the visibilities after using PIEFLAG, led to the removal of a few other small sections of RFI-affected data. In total, approximately 3$\%$ and 15$\%$ of the data were flagged out in the lower and higher bands respectively. Phase and amplitude fluctuations throughout the observing run were then corrected using the interleaved secondary calibrator data, and the amplitudes were scaled by bootstrapping to the primary calibrator. The data were then split by pointing position and each field was individually imaged, before mosaicing to form a master image, sensitivity and noise maps.

\subsection{Imaging}

\begin{figure*}
 \centering
 \includegraphics[width=0.98\linewidth,angle=0]{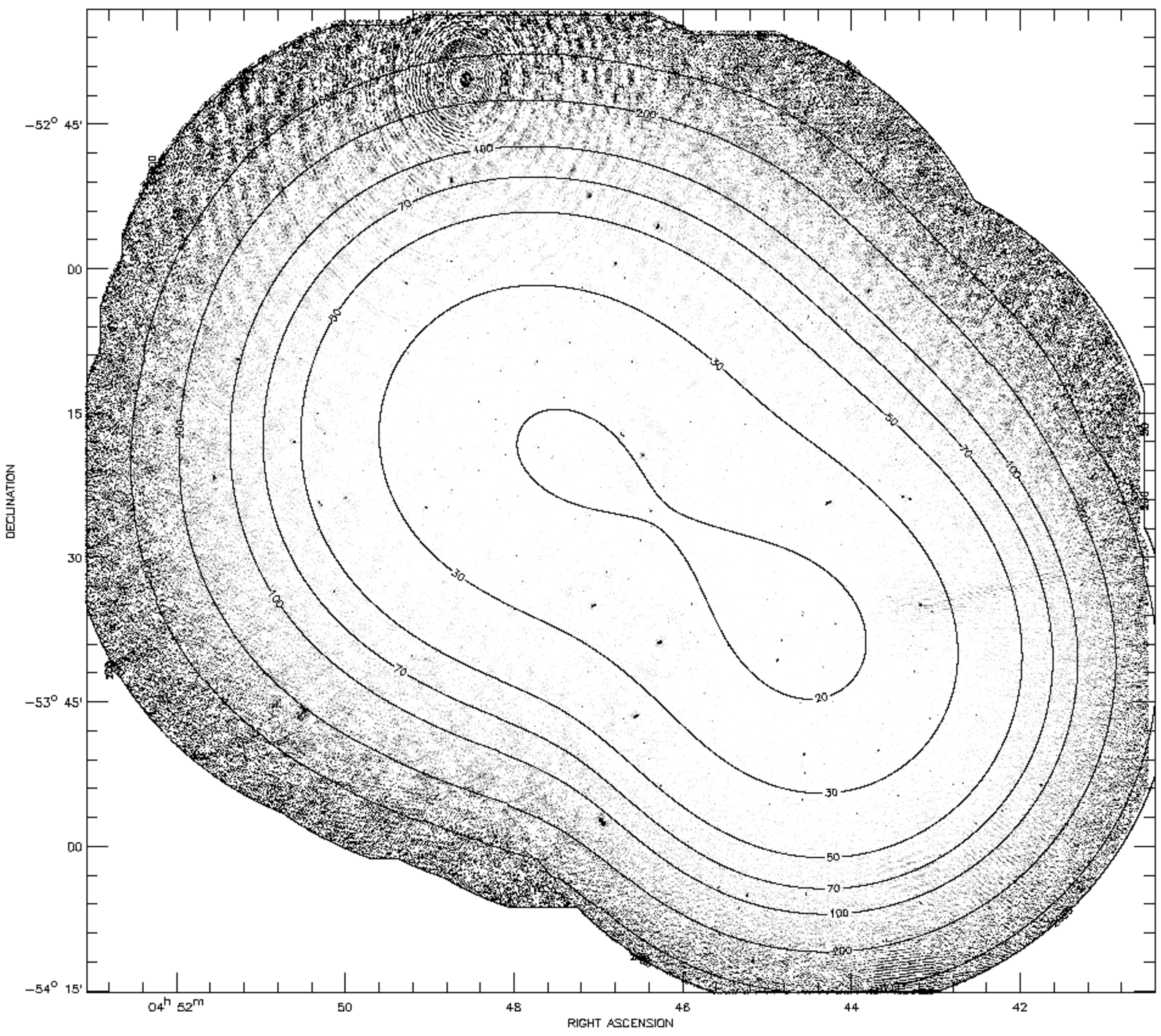}\\
 \caption{The central area of the ATCA 20 cm map, corrected for the primary beam of the antenna. The contours show the rms noise levels in $\mu$Jy beam$^{-1}$ estimated locally from the noise map by binning the data into 40$\times$40 pixel regions.}
 \label{surveyarea}
\end{figure*}

The data for each of the target pointings were imaged separately using uniform weighting and gridded to a pixel size of 2.0\arcsec ~to a common reference frame (to minimise geometrical issues in the mosaicing process).  The twenty-five 8 MHz wide frequency channels across the ATCA passband were reduced using ${\it MIRIAD}$'s implementation of multi-frequency clean, MFCLEAN, which accounts for variation in the spectral index  of the calibration sources across the observed bandwidth. After a first iteration of MFCLEAN, model components with flux densities $\geq$ 1 mJy beam$^{-1}$  were used to phase self-calibrate, and to correct residual phase errors. The data were then re-imaged and CLEANED for 5000 iterations, at which point the sidelobes of strong sources were generally found to be comparable with the thermal noise, except for a few cases adjacent to bright sources. The individual pointings were then mosaiced together using the ${\it MIRIAD}$ task LINMOS, which additionally divides each image by a model of the primary beam attenuation, and uses a weighted average of positions contained in more than one pointing. As a result, pixels at the mosaic edges have a higher noise level. Regions beyond the point where the primary beam response drops below 50$\%$ (this occurs at a radius of ~35.06\arcmin ~from the centre of a pointing) were blanked, which resulted in a total survey area of 1.04 deg.$^2$ (to the limit of the half power beam width at the edges of the master image). The synthesised beam size in the final mosaiced image was 6.2\arcsec $\times$ 4.9\arcsec ~at a position angle of 0 degrees. The sensitivity varies across the image due to primary beam attenuation and the mosaicing strategy as shown in Figure \ref{surveyarea}, although the noise level achieved across the map is $\sim$35$\%$ higher than expected for a thermal noise limited survey, which is due to difficulties in removing the sidelobes of strong sources at the edge of the survey field. This is a well known situation that has previously been seen both for ATCA and WSRT radio surveys, and probably results from both the non-circularity of the telescope beam, and small movements of the primary beam on the sky caused by random single dish pointing errors (i.e. due to wind/thermal loading) that cause the intensity of bright sources near the edge of the primary beam to vary significantly during an integration, making it difficult to efficiently CLEAN those areas.

\section{Source Component Catalogue}
\label{detection}
\begin{figure}
 \centering
 \includegraphics[width=0.98\linewidth]{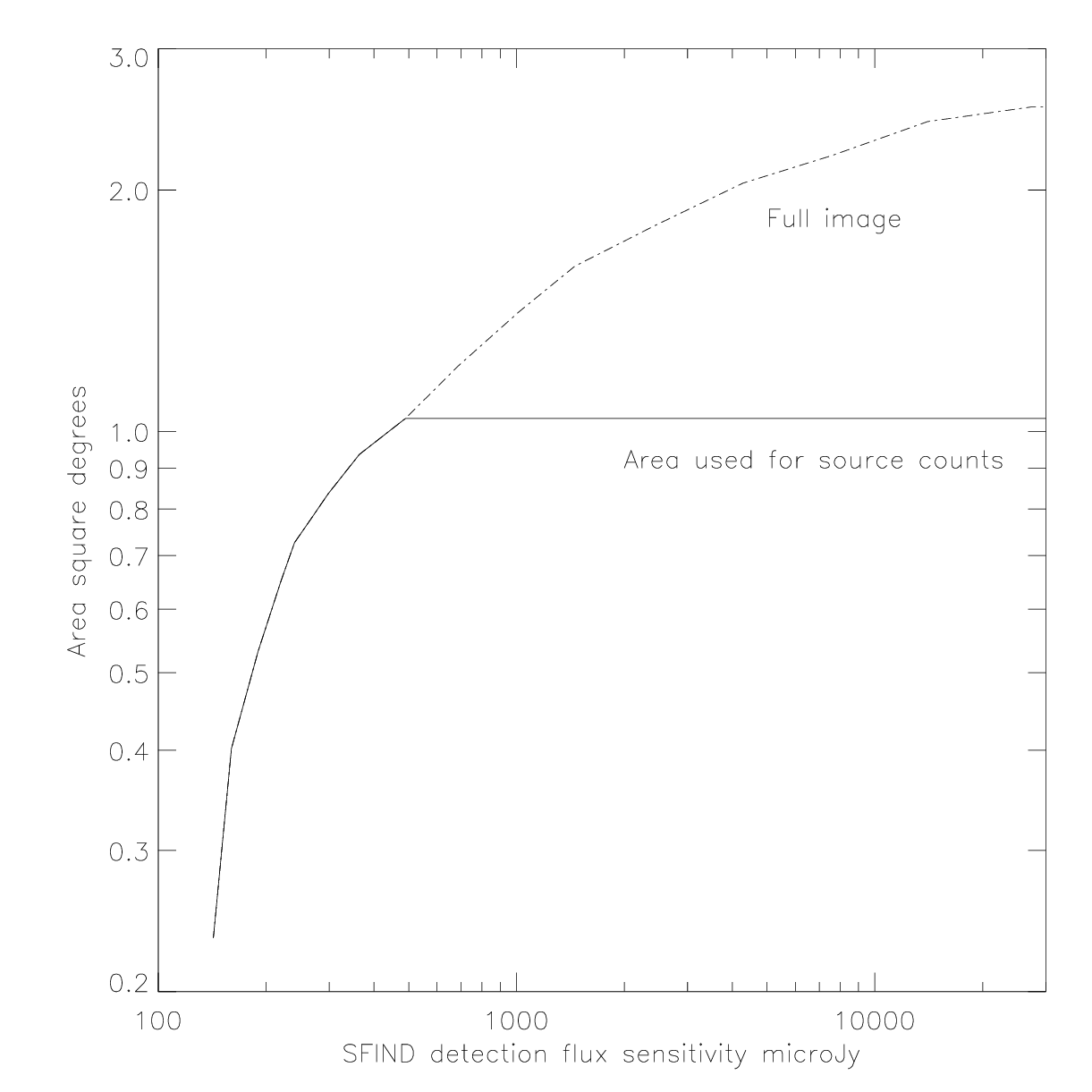}\\
 \caption{The horizontal axis shows the ${\it SFIND}$ detection threshold as a function of areal coverage. The area used for the differential source count estimation in Section \ref{sect:diffsourcecounts} is shown as a solid line, and has a maximum value of 1.04 degree$^2$, whereas that of the full image (whose radio components are listed in Table \ref{sourcecatalogueshort}) is indicated by the dot-dash line and has a maximum value of 2.55 degree$^2$.}
 \label{area}
\end{figure}

The mosaiced region achieves wide-field coverage and good sensitivity at the price of having an unavoidably non-uniform noise distribution. Statistical characterisation of the completeness of detection at various flux levels is therefore a  complex procedure that requires accounting for the observing time, mosaic overlap, and primary beam attenuation. Our source detection was made using locally determined noise levels derived from the noise map (Figure \ref{area}) - an approach that has already been used in other studies to improve the efficacy of their source detection catalogues (e.g. Hopkins \al 1998, Morganti \al 2004, Paper 1, and the associated NEP component catalogue presented in White \al 2010b).

The component catalogue in this paper was built using the ${\it MIRIAD}$ task ${\it SFIND}$ in a similar way to that described in Paper 1. However, briefly ${\it SFIND}$ uses a statistical technique, the false discovery rate (FDR), which assigns a threshold based on an acceptable rate of false detections (Hopkins \al 2002). For the ATCA-ADFS data the approach of Hopkins \al (2002) was followed by adopting an FDR value of 2$\%$. The components identified by ${\it SFIND}$ were visually inspected to remove any obvious mis-identifications (e.g. a few residual sidelobe structures immediately adjacent to the brightest components in the mapped region). Comparison with independent catalogues derived using the ${\it MIRIAD}$ task ${\it IMSAD}$ (with a 7~$\sigma$ clip), and with one derived using ${\it SExtractor}$ (Bertin \&  Arnouts 1996) with a locally defined background ${\it rms}$ were almost identical with the ${\it SFIND}$ catalogue. Hopkins \al (1998) show that using ${\it SFIND}$ in this way provides a very robust estimate of the noise level above which there are almost no spurious positive candidates, with the completeness being robustly set by the choice of FDR, and the locally determined background noise level. An understanding of source confusion, spurious components, sensitivity and completeness are important in any survey that is analysed to its limit, but as this becomes difficult to rigorously establish for mosaiced images with non-uniform noise properties of our mosaic and the fact that some but not all of the components are resolved, it was decided for the source counts analysis in Section \ref{sect:diffsourcecounts} to stop the calculation at the very conservative level of 200 $\mu$Jy, which corresponds to in excess of 10$\sigma$ signal to noise in the most sensitive parts of the mapped region.

A sample from the final component catalogue is presented in Table~ \ref{sourcecatalogueshort}, and the entire catalogue is included in the electronic on-line version of this paper.

\begin{table*}
\vspace{0pt}
\caption{The component catalogue (the full version is available as Supplementary Material in the on-line version of this article). The component parameters listed in the catalog are: (1) a short form running number (components that are believed to be parts of multi-component sources are listed with a ${\bf ^{\dagger}}$ sign next to the running number (for example 47${\bf ^{\dagger}}$), with more details about these multi-component sources being presented in Table \ref{doublecatalogueshort}, (2) the component name, referred to in this paper as ATCA-ADFS followed by the RA/Dec encoding (e.g. ATCA-ADFS J045243-533127), (3,4) the component Right Ascension and Declination (J2000) referenced from the self-calibrated reference frame, (5,6) the RA and Dec errors in arc seconds, (7,8) the peak flux density, S$_{\rm peak}$, and its associated rms error, (9,10) the integrated flux densities, S$_{\rm total}$ and their  associated errors, (11, 12, 13) the size along the major and minor axes of the fitted Gaussian component profile and its orientation (the major and minor axes refer to the full width at half maximum component size deconvolved from the synthesised beam, and position angle was measured east of north. Component sizes are shown in columns 11 or 12 only for the cases where S$_{\rm total}$/S$_{\rm peak}$ $\ge$ 1.3, as an indicator of a resolved component. Components where S$_{\rm total}$/S$_{\rm peak}$ $<$ 1.3 were considered to be unresolved, and therefore component sizes are not individually reported for these here. All components were additionally checked visually to mitigate against artefacts that might have slipped through the various checks.}
\begin{scriptsize}
\fontsize{8}{10}\selectfont
\begin{tabular}{l l r r r r r r r r r r r}
\hline
\multicolumn{1}{l}{No} & \multicolumn{1}{c}{Component name} & \multicolumn{1}{c}{RA} & \multicolumn{1}{c}{DEC} & \multicolumn{1}{c}{$\Delta$RA} & \multicolumn{1}{c}{$\Delta$DEC} & \multicolumn{1}{c}{S$_{\rm peak}$} & \multicolumn{1}{c}{$\Delta$S$_{\rm peak}$} & \multicolumn{1}{c}{S$_{\rm total}$} & \multicolumn{1}{c}{$\Delta$S$_{\rm total}$} & \multicolumn{1}{c}{$\theta_{maj}$} & \multicolumn{1}{c}{$\theta_{min}$} & \multicolumn{1}{c}{$PA$}\\ 
 \multicolumn{1}{l}{} &         & \multicolumn{1}{c}{h:m:s.s} & \multicolumn{1}{c}{d:m:s.s} & \multicolumn{1}{c}{${\prime\prime}$} & \multicolumn{1}{c}{${\prime\prime}$} & \multicolumn{1}{c}{mJy} & \multicolumn{1}{c}{mJy} & \multicolumn{1}{c}{mJy}  & \multicolumn{1}{c}{mJy}  & \multicolumn{1}{c}{$^{\prime\prime}$} &  \multicolumn{1}{c}{$^{\prime\prime}$} &  \multicolumn{1}{c}{$\ensuremath{^\circ}\,$}\\
  \multicolumn{1}{l}{} &         & \multicolumn{1}{c}{} & \multicolumn{1}{c}{} & \multicolumn{1}{c}{} & \multicolumn{1}{c}{} & \multicolumn{1}{c}{beam$^{-1}$} & \multicolumn{1}{c}{beam$^{-1}$} & \multicolumn{1}{c}{}  & \multicolumn{1}{c}{}  & \multicolumn{1}{c}{} &  \multicolumn{1}{c}{} &  \multicolumn{1}{c}{}\\
\multicolumn{1}{l}{(1)} & \multicolumn{1}{c}{(2)} & \multicolumn{1}{c}{(3)} & \multicolumn{1}{c}{(4)} &  \multicolumn{1}{c}{(5)} & \multicolumn{1}{c}{(6)} & \multicolumn{1}{c}{(7)} & \multicolumn{1}{c}{(8)} & \multicolumn{1}{c}{(9)} & \multicolumn{1}{c}{(10)} & \multicolumn{1}{c}{(11)}  & \multicolumn{1}{c}{(12)} & \multicolumn{1}{c}{(13)} \\
\hline
 1 &  ATCA-ADFS J044041-534043 & 4:40:41.5 & -53:40:43.5 & 0.14 & 0.03 &   2.951 &   0.286 &   7.392 &   0.368 &   12.1 &    2.7 & -84.0\\
 2 &  ATCA-ADFS J044116-532554 & 4:41:16.4 & -53:25:54.0 & 0.00 & 0.00 &   1.498 &   0.153 &   4.427 &   0.206 &    8.7 &    6.6 & -48.1\\
 3 &  ATCA-ADFS J044116-531845 & 4:41:16.9 & -53:18:45.8 & 0.07 & 0.31 &   1.568 &   0.260 &   5.310 &   0.317 &   13.5 &    3.5 &  -9.3\\
 4 &  ATCA-ADFS J044117-531853 & 4:41:17.9 & -53:18:53.5 & 0.03 & 0.02 &   1.866 &   0.260 &   2.632 &   0.274 &    6.7 &  & -40.6\\
 5 &  ATCA-ADFS J044120-533214 & 4:41:20.1 & -53:32:14.9 & 0.03 & 0.02 &   1.603 &   0.116 &   5.893 &   0.137 &   14.1 &    4.9 & -53.8\\
 6 &  ATCA-ADFS J044120-531626 & 4:41:20.7 & -53:16:26.3 & 0.79 & 0.16 &   1.626 &   0.260 &   6.553 &   0.359 &   14.8 &    5.8 & -87.5\\
 7 &  ATCA-ADFS J044121-531637 & 4:41:21.9 & -53:16:37.9 & 0.17 & 0.05 &   1.966 &   0.260 &   7.431 &   0.361 &   12.6 &    6.6 & -88.1\\
 8 &  ATCA-ADFS J044122-534349 & 4:41:22.0 & -53:43:49.1 & 0.01 & 0.01 &   0.686 &   0.096 &   1.051 &   0.100 &  &  & \\
 9 &  ATCA-ADFS J044123-534402 & 4:41:23.8 & -53:44:02.8 & 0.07 & 0.01 &   0.932 &   0.096 &   2.413 &   0.121 &   10.3 &    4.4 & -86.0\\
 10 &  ATCA-ADFS J044124-531600 & 4:41:24.1 & -53:16:00.1 & 0.32 & 0.13 &   1.677 &   0.260 &   4.353 &   0.340 &    8.6 &    5.5 &  86.1\\
 11 &  ATCA-ADFS J044141-533707 & 4:41:41.9 & -53:37:07.6 & 0.00 & 0.00 &   6.011 &   0.116 &  11.775 &   0.178 &    7.5 &    3.6 & -80.4\\
 12 &  ATCA-ADFS J044145-535304 & 4:41:45.5 & -53:53:04.4 & 0.00 & 0.00 &  11.728 &   0.187 &  23.327 &   0.210 &    8.7 &    2.0 &  59.2\\
 13 &  ATCA-ADFS J044156-531452 & 4:41:56.5 & -53:14:52.4 & 0.00 & 0.01 &   4.390 &   0.507 &  10.187 &   0.521 &    9.8 &    2.5 & -38.7\\
 14 &  ATCA-ADFS J044203-534302 & 4:42:03.9 & -53:43:02.1 & 0.00 & 0.00 &   0.639 &   0.050 &   0.726 &   0.050 &  &  & \\
 15 &  ATCA-ADFS J044205-533253 & 4:42:05.0 & -53:32:53.3 & 0.25 & 0.02 &   0.356 &   0.071 &   1.363 &   0.074 &   20.9 &  & -79.4\\
 16 &  ATCA-ADFS J044208-535941 & 4:42:08.4 & -53:59:41.2 & 0.41 & 0.04 &   0.584 &   0.080 &   0.923 &   0.095 &   12.2 &  & -73.9\\
 17 &  ATCA-ADFS J044212-531047 & 4:42:12.6 & -53:10:47.9 & 0.01 & 0.03 &   1.184 &   0.190 &   1.974 &   0.197 &    6.6 &  & -18.2\\
 18 &  ATCA-ADFS J044212-535551 & 4:42:12.6 & -53:55:51.6 & 0.01 & 0.01 &   1.374 &   0.080 &   2.291 &   0.087 &    7.8 &  &  47.6\\
 19 &  ATCA-ADFS J044212-530209 & 4:42:13.0 & -53:02:09.8 & 0.04 & 0.11 &   2.977 &   0.399 &   3.947 &   0.484 &    6.6 &  &   3.7\\
 20 &  ATCA-ADFS J044213-530802 & 4:42:13.6 & -53:08:02.5 & 0.06 & 0.11 &   1.166 &   0.190 &   2.322 &   0.216 &    9.4 &  & -23.0\\
\hline
 \end{tabular}
\end{scriptsize}
\label{sourcecatalogueshort}
\end{table*}


The positional accuracy listed in the Table \ref{sourcecatalogueshort} is relative to the self-calibrated and bootstrapped reference frame described in Section \ref{section:observations}. Other effects that bias the positions or sizes of sources in radio surveys have already been presented in Paper 1, to which the reader is referred. An estimate of component dimensions calculated by deconvolving the measured sizes from the synthesised beam is also presented, with Table \ref{sourcecatalogueshort} reporting only those more than double the synthesised beam size.

\subsection{Component extraction}

\begin{figure*}
 \centering
 \includegraphics[width=1.0\linewidth,angle=0]{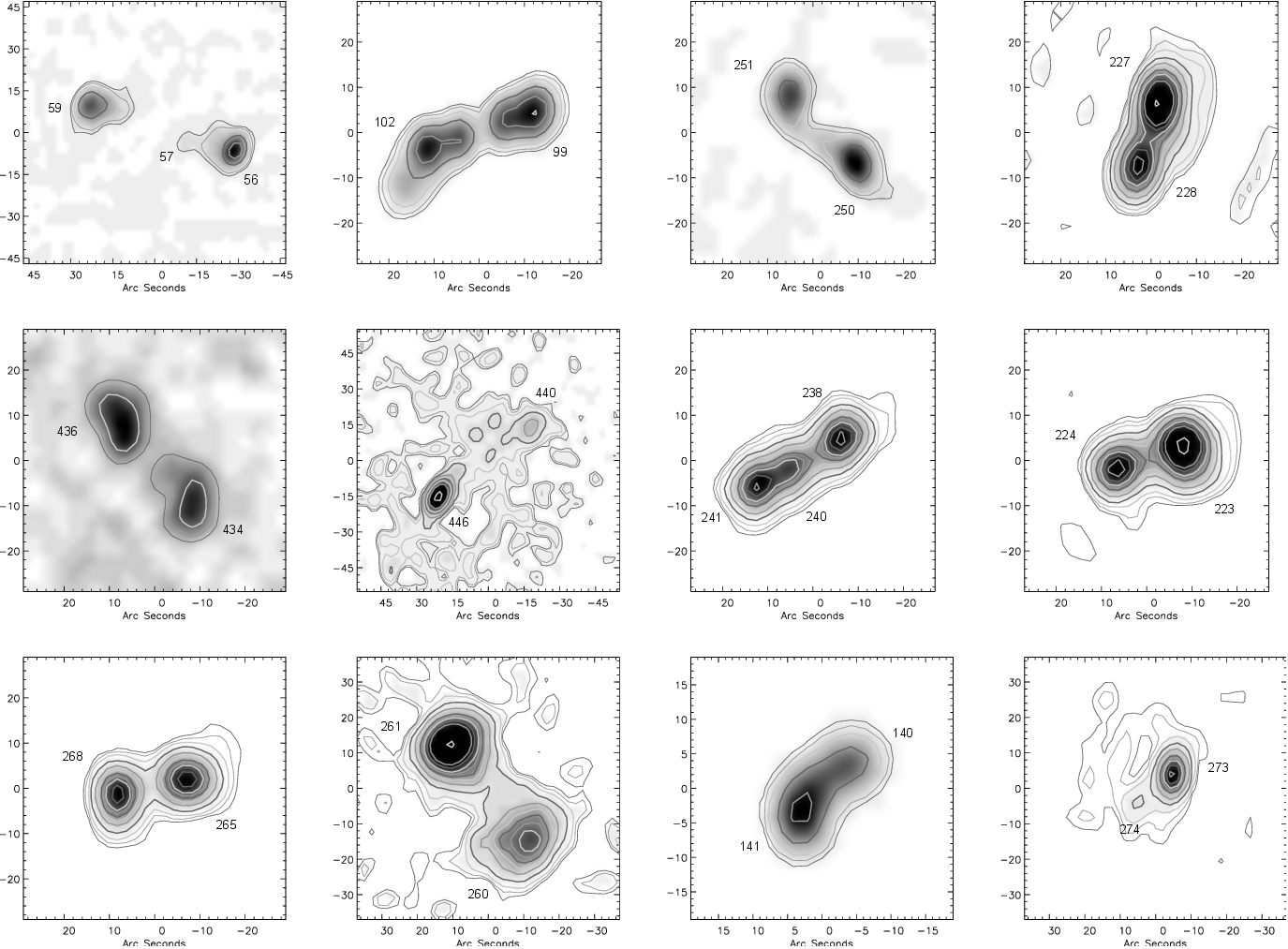}\\
 \caption{Regions showing complex or extended structure. The vertical scale is Declination. The contours are at 0.0001, 0.0003, 0.0005, 0.001, 0.003, 0.006, 0.012, 0.024, 0.048 and 0.096 Jy beam$^{-1}$ respectively. The Right Ascension/Declination scales can de derived using the component locations in Table \ref{sourcecatalogueshort}.}
 \label{ATCAresolvedsources}
\end{figure*}

In the terminology of this paper, a radio component is described as a region of radio emission represented by a Gaussian shaped object in the map. Close radio doubles are represented by two Gaussians and are deemed to consist of two components, which make up a single source. A selection of radio sources with multiple components is shown in Figure \ref{ATCAresolvedsources}.

\subsection{Complex sources}

\begin{figure}
 \includegraphics[width=1.0\linewidth]{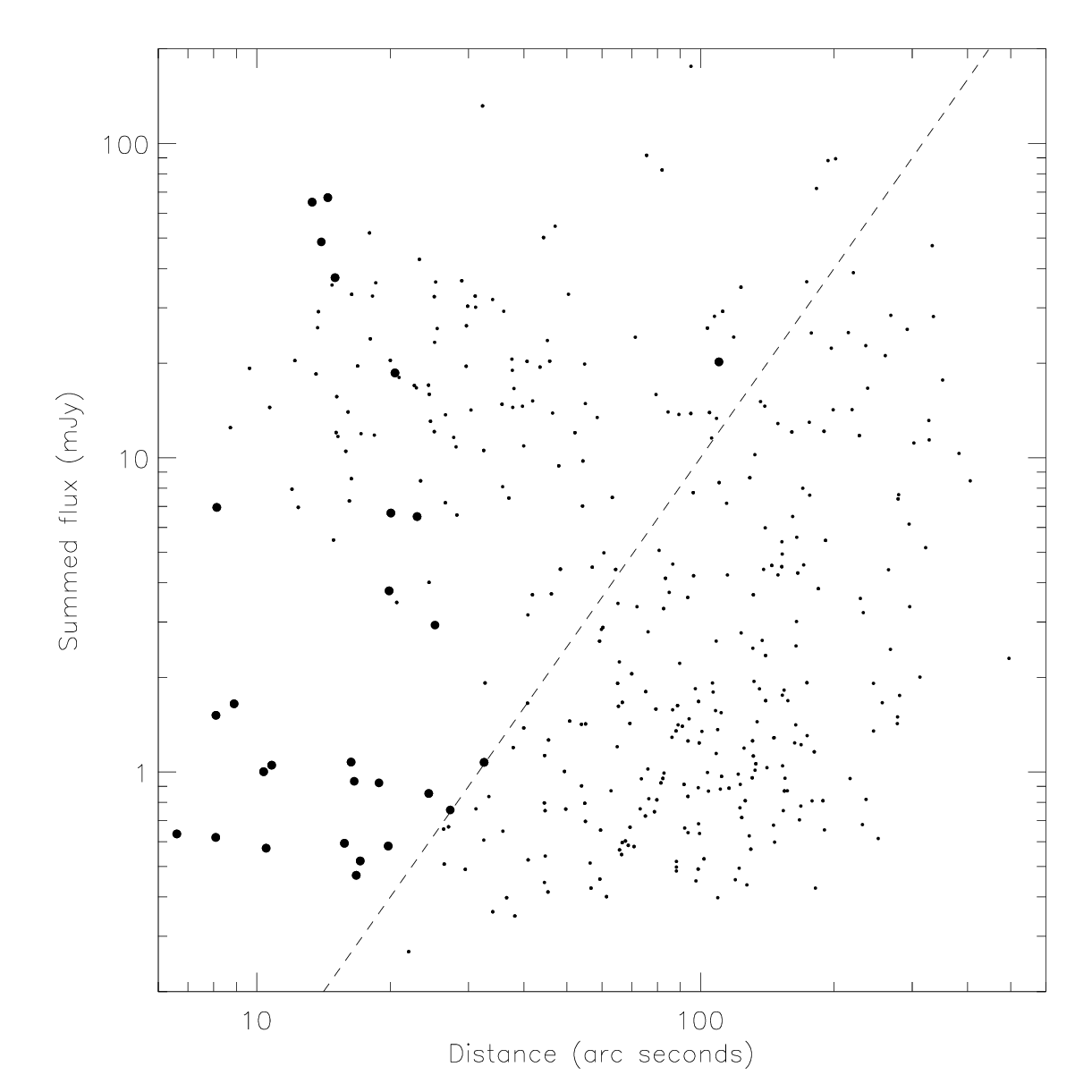}\\
 \caption{This Figure shows the sum of the flux densities of the nearest neighbours between components in the detection catalogue. Following Magliocchetti \al (1998) points to the left of the dashed line are possible double sources. The likelihood that two sources in a pair are related is further constrained (Magliocchetti \al 1998) by requiring that the fluxes of the two components $f_1$ and $f_2$ should be in the range 0.25 $\le$ $f_1 / f_2$ $\le$ 4. Sources in the Figure whose components satisfying this additional criterion are shown as bold circles.}
 \label{nearestneighbour}
\end{figure}

Radio sources are often made up of multiple components, as seen in Figure \ref{ATCAresolvedsources}. The source counts need to be corrected for the multi-component sorces, so that the fluxes of physically related components are summed together, rather than being treated as separate sources. Magliocchetti \al (1998) have proposed criteria to identify the double and compact source populations, by plotting the separation of the nearest neighbour of a component against the summed flux of the two components, and selecting components where the ratio of their fluxes, $f_1$ and $f_2$ is in the range 0.25 $\le$ $f_1 / f_2$ $\le$ 4. In Figure \ref{nearestneighbour} the sum of the fluxes of nearest neighbours are plotted against their separation.

The dashed line marks the boundary satisfying the separation criterion defined by Magliocchetti \al (1998):

\begin{equation}
\theta  = 100\left[ {\frac{{S_{\rm total} ({\rm mJy}) }}{{10}}} \right]^{0.5}
\end{equation}

where $\theta$ is in arc seconds. Therefore 53 radio sources in the present survey (i.e. 10$\%$ of the 530 catalogued entries) should be considered to be a part of double or multiple sources according to the Magliocchetti \al (1998) criterion, and this will be taken account of in the source counts discussed later. These components, and their suggested associations are listed in Table \ref{doublecatalogueshort}.

\begin{table}
\vspace{0pt}
\caption{The multi-component source catalogue for components satisfying the Magliocchetti \al (1998) criterion. The proposed multi-component sources listed in this Table are: (1) the components identified according to their Running Numbers in the main catalogue, (2,3) mean Right Ascension and Declination (J2000) taken and the average of the positions of the individual components, (4) the distance between the components (rounded up to the nearest arc second), (5) the sum of the total flux density of the individual components, (6) the error on this, estimated by adding the total flux errors in quadrature.}
\begin{scriptsize}
\fontsize{8}{10}\selectfont
\begin{tabular}{c l l c r r }
\hline
\multicolumn{1}{l}{Components} & \multicolumn{1}{c}{RA} & \multicolumn{1}{c}{Dec} & \multicolumn{1}{c}{Dist} & \multicolumn{1}{c}{S$_{c}$} & \multicolumn{1}{c}{$\Delta$S$_{c}$}\\
\multicolumn{1}{l}{} & \multicolumn{1}{c}{h:m:s.s} & \multicolumn{1}{c}{d:m:s.s} & \multicolumn{1}{c}{$^{\prime\prime}$} & \multicolumn{1}{c}{mJy} & \multicolumn{1}{c}{mJy}\\
\multicolumn{1}{l}{(1)} & \multicolumn{1}{c}{(2)} & \multicolumn{1}{c}{(3)} & \multicolumn{1}{c}{(4)} &  \multicolumn{1}{c}{(5)} & \multicolumn{1}{c}{(6)}\\
\hline

47+49		& 4:43:09.2& -53:39:30.0	& 9 & 1.651 & 0.029\\
65+67		& 4:43:38.8& -53:47:24.1	& 17 & 0.521 & 0.026\\
69+70		&4:43:43.3&-53:22:18.1	&11&1.050&0.034\\
99+102		&4:44:18.9&-53:24:38.8	&20&18.617&0.241\\
122+123		&4:44:35.9&-53:36:01.4	&8&1.515&0.031\\
140+141		&4:44:53.1&-53:41:00.4	&8&6.949&0.058\\
166+169		&4:45:19.3&-53:12:26.2	&16&1.075&0.028\\
171+172+175	&4:45:22.6&-53:09:09.4&20&1.450&0.021\\
179+180		&4:45:25.8&-53:06:08.2 &16&0.593&0.042\\
212+213		&4:46:03.1&-53:44:41.6	&17&0.469&0.026\\
223+224		&4:46:16.6&-53:39:21.9	&15&67.264&0.431\\
227+228		&4:46:20.4&-52:56:23.8	&13&65.058&0.445\\
238+240+241 &4:46:32.6	&-53:47:00.4	&18&71.759&0.172\\
250+251		&4:46:43.7&-53:17:59.8	&20&6.669&0.062\\
248+253		&4:46:46.0&-53:01:00.5	&110&20.188&0.141\\
265+268		&4:47:03.2&-53:35:39.4	&15&37.415&0.181\\
284+286		&4:47:16.2&-52:59:27.8	&17&0.935&0.399\\
305+306		&4:47:32.2&-53:20:22.8	&10&1.102&0.020\\
318+320		&4:47:44.0&-53:10:37.4	&20&3.771&0.052\\
336+337		&4:48:02.5&-53:26:14.4	&7&0.635&0.031\\
371+373		&4:48:40.6&-52:59:31.5	&33&1.073&0.399\\
378+379+380	&4:48:44.9&-53:00:04.8	&30&3.712&0.043\\
410+411		&4:49:28.5&-53:26:05.0	&11&0.572&0.026\\
425+428		&4:49:53.7&-53:12:02.8	&28&0.854&0.036\\
434+436		&4:50:14.7&-53:25:08.6	&23&6.499&0.049\\
\hline
 \end{tabular}
\end{scriptsize}
\label{doublecatalogueshort}
\end{table}

\subsection{Flux density and positional accuracy}
 \label{bright_galaxy_offsets}
\begin{figure}
 \centering
 \includegraphics[width=1.2\linewidth]{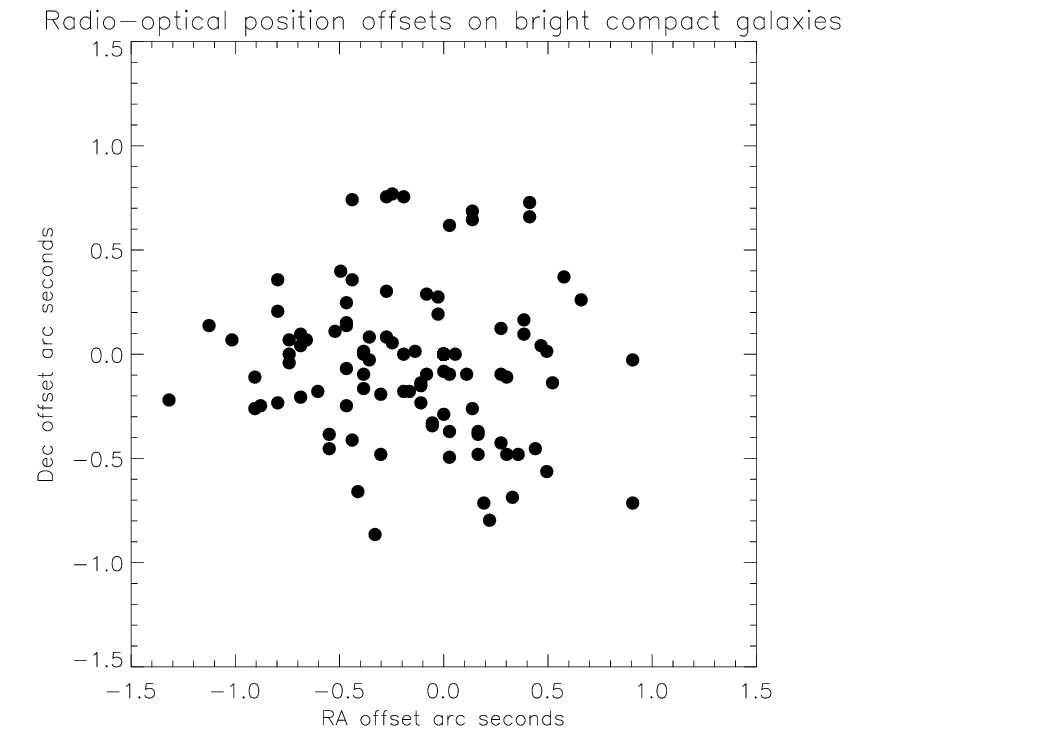}\\
 \caption{Positional offsets between bright (R-magnitude $\le$ 22 mag) galaxies from the CTIO MOSAIC-II images and the DENIS database, and a sample of ATCA radio components detected at $\ge$ 10$\sigma$ levels in Table \ref{sourcecatalogueshort}.}
\label{Fig:radio_optical}
\end{figure}

The flux density and positional accuracy are presented in Table \ref{sourcecatalogueshort}, and the method for calculating the positional accuracy are described in Hopkins \al (2002), and the intensity scales are derived and fully described in Equations 1--5 of Hopkins \al (2003). Since the methods for measuring the positional and intensity scale accuracy form part of the methodology of the ${\it SFIND}$ technique, the reader is referred to the papers presenting this technique, rather than repeating them here. However, to check the positional accuracy, the ATCA data were cross correlated against the SUMSS survey (Mauch \al 2003), where 8 of the bright ATCA sources were found to be within 10\arcsec ~of a SUMSS source (the SUMSS half-power beam width is 45\arcsec~$\times$~57\arcsec). After eliminating three components which are resolved and appear as double radio sources in the ATCA data, the average offset between the positions in the two catalogues (ATCA-SUMSS) was ($\Delta$RA,$\Delta$Dec) = (+0.43\arcsec~$\pm$~2.31\arcsec, -2.57\arcsec~$\pm$2.56\arcsec), which are consistent with the absolute and systematic errors reported in the SUMSS Catalogue. The ATCA component catalogue was also cross-correlated with the positions of bright compact optical galaxies from our CTIO MOSAIC-II survey (see Table \ref{ancillary}), which was astrometrically referenced against HST guide stars, and sources from the DENIS database. The mean of the offsets to the 166 bright galaxies shown in Figure \ref{Fig:radio_optical} was $\Delta$ RA = -0.16\arcsec$~\pm$~0.37\arcsec ~and  $\Delta$ Dec = -0.05\arcsec~$\pm$~0.46\arcsec, which is also consistent with the SUMSS result. 

\subsection{Summary of flux density corrections for systematic effects}

There are two main systematic effects which have been taken into account to estimate the ATCA flux densities, specifically clean bias and bandwidth smearing effects. Bandwidth smearing is the radio analog of optical chromatic aberration, resulting from the finite width of the receiver channels compared to the observing frequency. It reduces the peak flux density of a source while correspondingly increasing, or blurring, the source size in the radial direction such that the total integrated flux density is conserved, but the peak flux is reduced. 

From Condon \al (1998) the reduction in the peak flux from a compact radio source as a result of bandwidth smearing is given by:

\begin{equation}
\frac{{S_{peak} }}{{S_{peak}^0 }} \cong \frac{1}{{\sqrt {1 + \left[ {\frac{{2\,\ln \,2}}{3}} \right]\left[ {\left( {\frac{{{\Delta\nu }}}{\nu}} \right)\left( {\frac{d}{{\theta _b }}} \right)} \right]^2 } }}
\end{equation}

where ${S_{peak}}$ and  ${{S_{peak}^0 }}$ refer to the off-axis peak flux and the peak flux at the centre of axis of the primary beam,$\Delta\nu$ and  ${\nu}$ are the bandwidth and observing frequency respectively, ${\it d}$ is the off-axis distance, and ${\theta _b }$ is the synthesised beamwidth. Prandoni \al (2000a) have made a detailed study of this for the ATCA telescope, finding similar behaviour.

\begin{figure}
 \centering
 \includegraphics[width=1.0\linewidth]{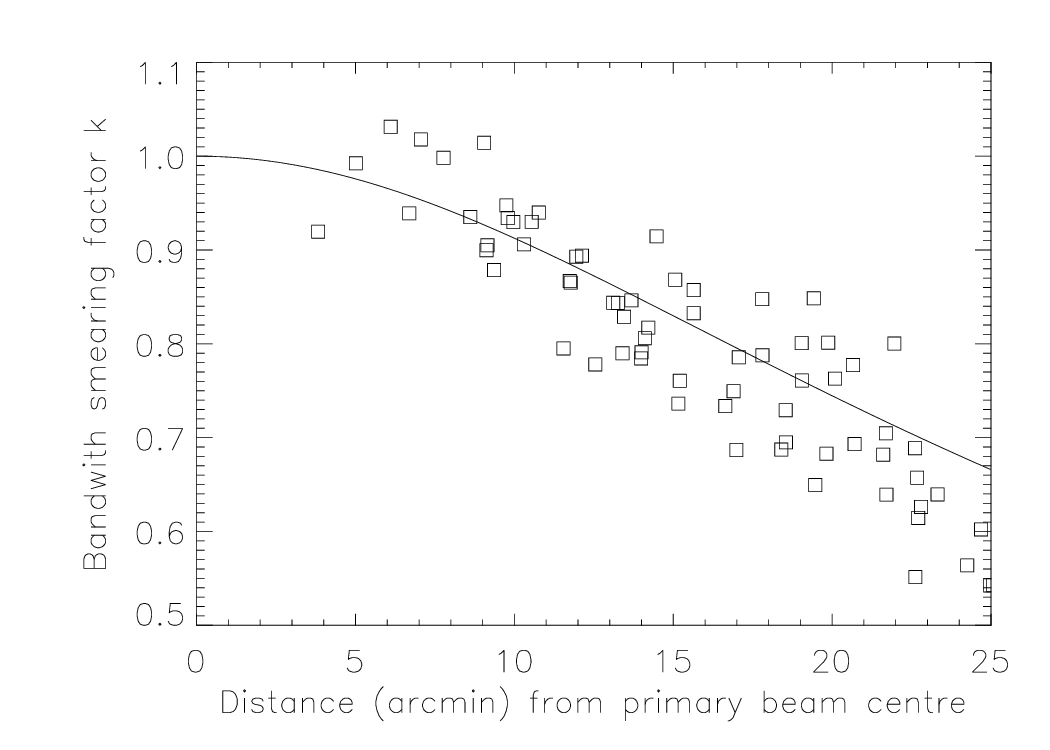}\\
 \caption{Variation of the smearing factor with distance from the centre of an individual field for a sample of components in common to many of the individual fields. The solid curve is the expected theoretical curve from Condon \al (1998). which matches closely to that shown by Huynh \al (2005) in their ATCA observation of the Hubble Deep Field South, and the squares show the .}
\label{Fig:smearing}
\end{figure}

Fortunately, the closely spaced mosaicing strategy used for the ATCA SEP observations allows the smearing effect to be measured directly, by monitoring peak and integrated flux densities of four bright compact sources that were present in virtually every one of the observed fields, but at different distances from the centre of the beam. Figure \ref{Fig:smearing} shows the measured smearing factor ${\it k}$, which we define as the ratio of the peak flux of a compact source normalised to that which it has when at the centre of a beam, as a function of distance from the beam centre. 

This Figure shows that the experimental data points are reasonably well fit by the theoretical relationship of Condon \al (1998), which is overlaid as a solid line on Figure \ref{Fig:smearing}. This correction was taken into account when estimating the peak fluxes listed in Table \ref{sourcecatalogueshort}.

It is well known that as well as needing to consider this effect for single pointings,  large mosaiced fields and chromatic aberration, it can also act to reduce point source fluxes in a complex way (e.g. Bondi \al 2008, White \al 2010). {\bf We have empirically examined the effect of bandwidth smearing on our mosaiced data by following the approach adopted by Bondi \al (2008) to compare the fluxes of bright sources observed close to the centres of individual pointings, with their fluxes determined after mosaicing together to form a merged image. Although the bandwidth smearing can be accounted for using the above equation from Condon \al (1998), as Bondi \al (2008) discuss, the contribution of this to measurements of the peak fluxes in radio surveys is more difficult to rigorously quantify for mosaiced data, where the smearing would need to be modelled with a more complicated function that represents the spacing pattern of the individual pointings. Due to the difficulty in rigorously calculating this, we have therefore followed the Bondi \al (2008) approach to estimate the most probable reduction to the peak flux densities, as this correction will slightly modify estimates of the estimated source sizes.

Therefore, we ran the same procedure that was used to produce the final radio catalogue, on each of the individual pointings. For the strongest unresolved sources ($\ga$ 1 mJy) the peak and total flux densities measured from the final mosaiced image, were compared with the corresponding peak and total flux densities from the individual pointings, using sources that were no further than 5$^{\prime}$ away from an individual pointing (this is consistent with the Bondi \al 2008 approach for the VLA, which is supported for ATCA from our own results shown in Fig. \ref{smearing_2}). The total flux densities of each source in the mosaic were in good agreement (the median value was measured to be 1.01 with an rms dispersion of 0.02), as would be expected for complete recovery of the flux. However, for components whose peak fluxes are affected by bandwidth smearing, the peak fluxes could be underestimated in the final mosaic on average by up to 20$\%$. The peak fluxes listed in Table \ref{sourcecatalogueshort} have all been corrected for this, according to the centre of the mosaiced image of Right Ascension (J2000) = 4$^h$ 46$^m$ 46$^s$.5, Declination (J2000) -53$^{\circ}$ 24$^{\prime}$ 59$^{\prime\prime}$.0. 

\begin{figure}
 \centering
 \includegraphics[width=0.76\linewidth,angle=90]{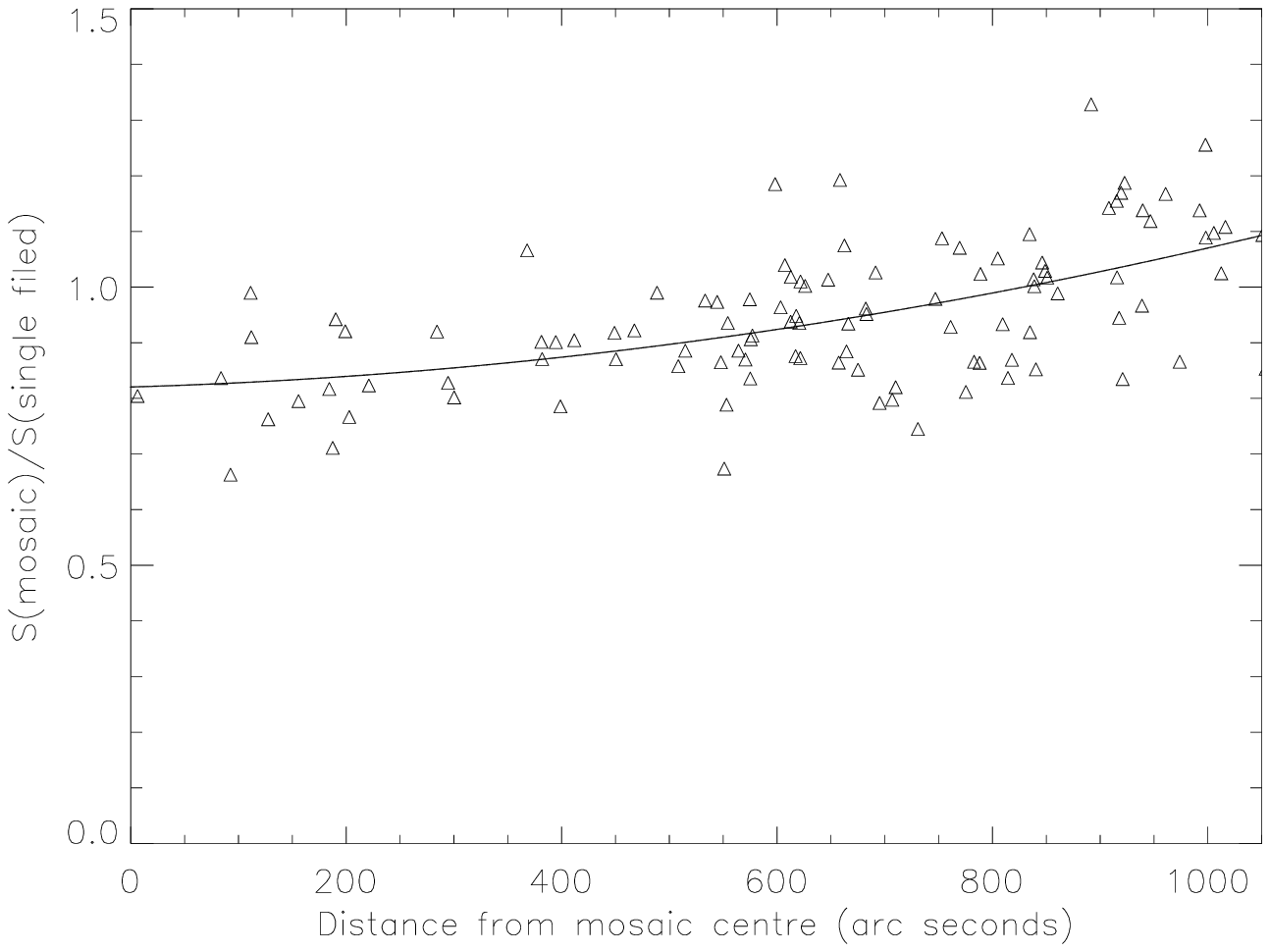}\\
 \caption{Ratio between the peak flux densities in the final mosaic and in the individual pointing where the source is within 5$^{\prime}$ from the centre vs. the radial distance from the centre of the final mosaic. Only compact sources with flux densities greater than 1 mJy beam$^{-1}$ are plotted. The fitted curve corresponds to a second order polynomial fit to the data represented by the relationship S(mosaic)/S(single field) = 0.82+5.64 10$^{-5}$~d$_{centre}$ + 1.93 10$^{-7}$~d$_{centre}$$^2$, whereS(mosaic)/S(single field) and d$_{centre}$, the distance in arc seconds from the centre of the mosaiced field, correspond to the vertical and horizontal axes.}
 \label{smearing_2}
\end{figure}

The other main effect that can influence fluxes is clean bias. Radio surveys, and in particular those consisting of short snapshot observations, have a tendency to be affected by the clean bias effect where the deconvolution process leads to a systematic underestimation of both the peak and total source fluxes. This is a consequence of the constraints on the cleaning algorithm due to sparse uv coverage (see Becker \al 1995, White \al 1997, Condon \al 1998), and has the effect of redistributing flux from point sources to noise peaks in the image, reducing the flux density of the real sources. As the amount of flux which is taken away from real sources is independent of the source flux densities, the fractional error this causes is most pronounced for weak sources. Prandoni \al (2000a, b) have shown that it is possible to mitigate clean bias if the CLEANing process is stopped well before the maximum residual flux has reached the theoretical noise level. Consequently the cleaning limit was set at 5 times the theoretical noise, to ensure that the clean bias does not significantly affect the source fluxes in the present survey (Garrett \al 2000). Gruppioni \al (1999) adopted a similar strategy in an ATCA survey of the ELAIS N1 field, and found the effect to be insignificant (less than 2.5$\%$) for the faintest sources (5$\sigma$ detections) but had no effect on sources brighter than 10$\sigma$ for similar numbers of CLEAN cycles as those performed on the present ATCA data. We therefore conclude that clean bias will have a negligible affect on the present data.
}

\section{Differential Counts}
\label{sect:diffsourcecounts}

\begin{figure}
 \centering
 \includegraphics[width=1.0\linewidth]{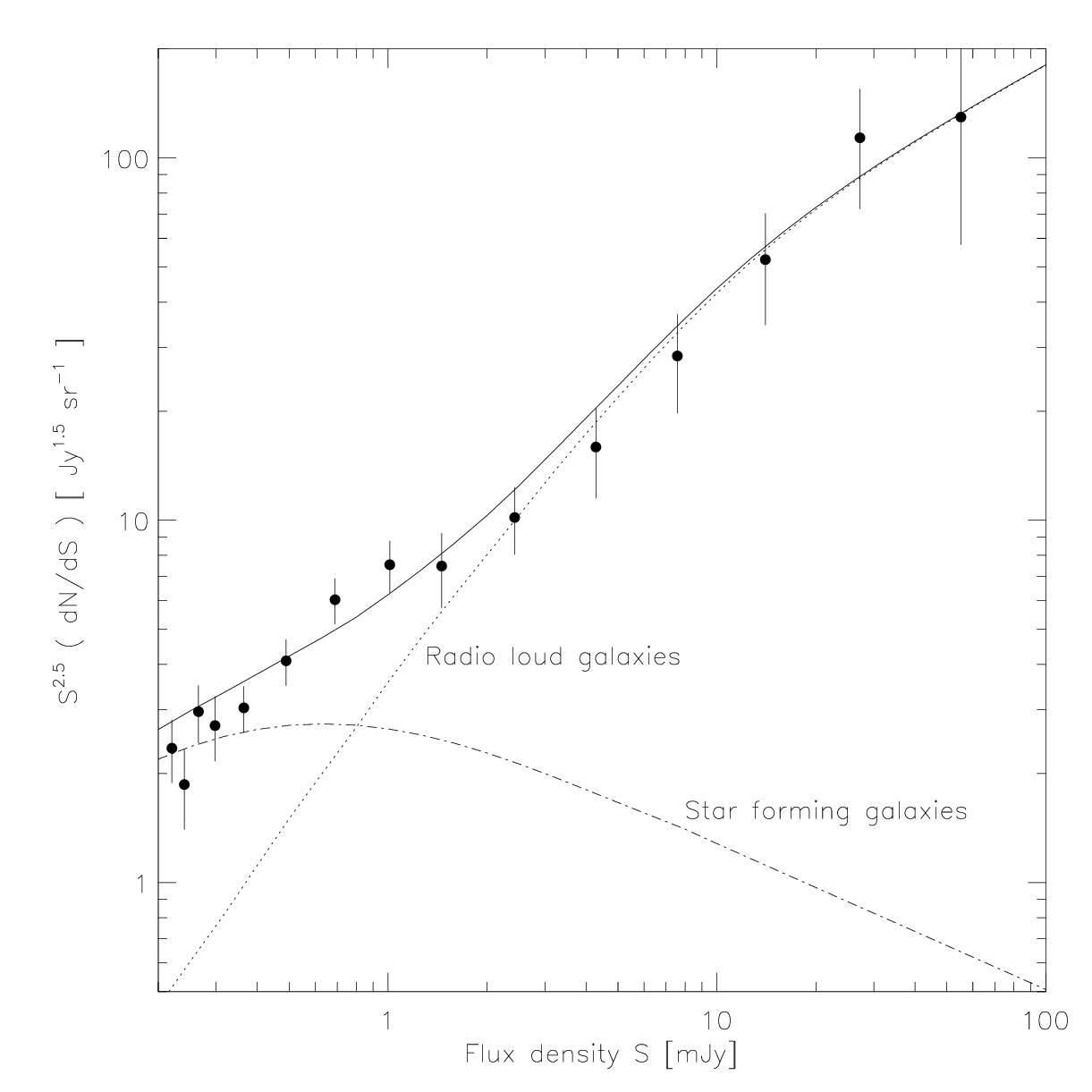}\\
 \caption{Differential counts determined from the AKARI ATCA-ADFS 20 cm deep field. The relationship for calculating the numbers in this plot and in Table \ref{Table:diff_source_numbers} is the same as that used by Kollgaard \al (1994).}
\label{ATCA_diff_counts}
\end{figure}

In Figure \ref{ATCA_diff_counts} the differential radio source counts are shown from the ATCA-ADFS field, normalised to a static Euclidean universe (d$N/$d$S~S^{2.5}$ (sr$^{-1}$mJy$^{1.5}$)). These source counts are broadly consistent with previous results at 1.4 GHz (e.g. the compilation of Windhorst \al (1993), the PHOENIX Deep Survey (Hopkins \al 2003), and the shallow ${\it NEP}$ survey of Kollgaard \al (1994)).

The data from Figure \ref{ATCA_diff_counts} are given in Table \ref{Table:diff_source_numbers}, where the integrated flux bins and mean fluxes for each of the bin centres are listed in columns (1 and 2), the number of sources corrected for clean and resolution bias are shown in column (3), and the number of sources corrected for the area coverage and multi-component sources in Column 4, and in the final column (5) the differential source counts and their associated errors as defined by Kollgaard \al (1994) are listed.


\begin{table}
\caption{20 cm differential source counts for the ATCA-ADFS survey}
\begin{tabular}{c c c c c}
\hline
\multicolumn{1}{c}{flux bin} & \multicolumn{1}{c}{mean flux} & \multicolumn{1}{c}{N$_0$} & \multicolumn{1}{c}{N$_c$} & \multicolumn{1}{c}{${\rm dN/dS}$} \\ 
 \multicolumn{1}{c}{mJy} &    \multicolumn{1}{c}{mJy}     & \multicolumn{1}{c}{} & \multicolumn{1}{c}{} & \multicolumn{1}{c}{sr$^{-1}$ Jy$^{1.5}$} \\ 
\multicolumn{1}{c}{(1)} & \multicolumn{1}{c}{(2)} & \multicolumn{1}{c}{(3)} & \multicolumn{1}{c}{(4)} &  \multicolumn{1}{c}{(5)} \\
\hline
~~~0.21--0.23 & 0.22 & 16 & 25.45 & 2.35$\pm$0.47\\
~~~0.23--0.25 & 0.24 & 14 & 19.50 & 2.24$\pm$0.51\\
~~~0.25--0.28 & 0.265 & 23 & 30.30 & 2.96$\pm$0.54\\
~~~0.28--0.315 & 0.298 & 20 & 24.24 & 2.72$\pm$0.55\\
~~~0.315--0.413 & 0.364 & 51 & 48.01 & 3.11$\pm$0.45\\
~~~0.413--0.566 & 0.490 & 49 & 50.72 & 4.37$\pm$0.61\\
~~~0.566--0.813 & 0.690 & 49 & 50.72 & 6.30$\pm$0.88\\
~~~0.813--1.21 & 1.011 & 32 & 37.77 & 7.54$\pm$1.23\\
~~~1.21--1.70 & 1.455 & 11 & 18.36 & 7.47$\pm$1.75\\
~~~1.70--3.151 & 2.425 & 17 & 22.66 & 10.18$\pm$2.14\\
~~~3.151--5.416 & 4.283 & 10 & 12.95 & 15.93$\pm$4.43\\
~~~5.416--9.742 & 7.579 & 9 & 10.79 & 28.41$\pm$8.65\\
~~~9.742--18.33 & 14.036 & 4 & 8.63 & 52.41$\pm$17.84\\
~~18.33--36.08 & 27.205 & 7 & 7.55 & 113.7$\pm$41.35\\
~~36.08--74.32 & 55.200 & 2 & 3.24 & 129.7$\pm$72.07\\
\hline \hline
\end{tabular}
\label{Table:diff_source_numbers}
\end{table}

To model the observed source counts a two component model was used consisting of a classical bright radio loud population and a fainter star-forming population. It is well established that classical bright radio galaxies require strong evolution in order to fit the observed source counts at radio wavelengths (Longair 1966, Rowan-Robinson 1970). The source counts above 10 mJy are dominated by giant radio galaxies and QSOs (powered by accretion onto black holes, commonly joined together in the literature under the generic term AGN). Radio loud sources dominate the source counts down to levels of $\sim$1 mJy, however, at the sub-mJy level the normalised source counts flatten as a new population of faint radio sources emerge (Windhorst \al 1985). The dominance of starburst galaxies in the sub-mJy population is already well established (Gruppioni \al 2008), where the number of blue galaxies with star-forming spectral signatures is seen to increase strongly. Rowan-Robinson \al (1980, 1993), Hopkins \al (1998), and others have concluded that the source counts at these faintest levels require two populations, AGNs and starburst galaxies. This latter population can best be modelled as a dusty star-forming population, under the assumption that it is the higher redshift analogue of the IRAS star-forming population (Rowan-Robinson \al 1993, Pearson $\&$ Rowan-Robinson 1996). In this scenario, the radio emission originates from the non-thermal synchrotron emission from relativistic electrons accelerated by supernovae remnants in the host galaxies.

To represent the radio loud population the luminosity function of Dunlop $\&$ Peacock (1990) was used  (parameters in Table C3 in their paper) to model the local space density with an assumption that the population evolves in luminosity with increasing redshift. The  luminosity evolution follows a power law with redshift of $(1+z)^{3.0}$, broadly consistent with both optically and X-ray selected quasars (Boyle \al 1987). The spectrum of the radio loud population was obtained from Elvis, Lockman $\&$ Fassnacht (1994), assuming a steep radio spectrum source of ($S_{\nu}\propto \nu ^{-\alpha}$, $\alpha$=1).

To model the faint sub-mJy population we use the {\it IRAS}  60~$\mu$m luminosity function of Saunders \al (2000), with the parameters for the star-forming population, defined by warmer 100~$\mu$m~/~60~$\mu$m   {\it IRAS} colours, given in Pearson (2001, 2005), and Sedgwick \al (2012).

To convert the infrared luminosity function to radio wavelengths, we derive below the ratio of the 60~$\mu$m luminosity to the radio luminosity, from the well established correlation between the  far-IR and radio flux (e.g. Helou, Soifer \& Rowan-Robinson (1985), Yun, Reddy \& Condon (2001), Appleton \al (2004)). 

Helou \al (1985) defined this relation between the far-infrared flux, $FIR/{\rm  W m^{-2}}$ and the 1.4 GHz radio emission, $S_{1.4 GHz}/{\rm  W m^{-2}Hz^{-1}}$ in terms of the ${\it q}$ factor given by, 

\begin{equation}
q = \log \left( \frac{{FIR}}{{3.75 \times 10^{12} {\rm  Hz}}} \right) - \log \left( \frac{{S_{1.4~GHz}}}{{\rm  W m^{-2}~Hz^{-1}}} \right)
\label{eqn:q}
\end{equation}

The far-infrared flux defined by Condon (1991) in terms of the 60~$\mu$m and 100~$\mu$m emission can be written as,

\begin{equation}
\left( {\frac{{FIR}}{{{\rm W m}^{{\rm  - 2}} }}} \right) = 1.26 \times 10^{12} \left( {\frac{{2.58~S_{60} + S_{100}}}{{{\rm W m}^{{\rm  - 2}} \,{\rm Hz}^{{\rm  - 1}} }}} \right)
\end{equation}

where the spectrum between 60~$\mu$m and 100~$\mu$m is defined by a spectral index ${\alpha}$, 

\begin{equation}
\frac{{S_{100} }}{{S_{60} }} = \left( {\frac{{\nu _{100} }}{{\nu _{60} }}} \right)^{ - \alpha }  \Rightarrow S_{100}  = 1.67^\alpha  \,S_{60} 
\end{equation}

such that,

\begin{equation}
\left( {\frac{{FIR}}{{{\rm W m}^{{\rm  - 2}} }}} \right) = 1.26 \times 10^{12} \left( {\frac{{2.58 + 1.67^\alpha}}{{{\rm W m}^{{\rm  - 2}} \,{\rm Hz}^{{\rm  - 1}} }}} \right)S_{60} 
\end{equation}

substituting the above relation into Equation \ref{eqn:q}, assuming a value of $q$=2.3 (Condon 1991,1992) and a value of $\alpha$=2.7 (Hacking \al 1987), it is then easy to show that:

\begin{equation}
S_{60}  \approx \,{\rm 90}\,S_{{\rm 1}{\rm .4}\,\,{\rm GHz}} 
\end{equation}

To convert the infrared luminosity function to radio wavelengths we adopt the above $S_{60~{\mu{\rm m}}}/S_{1.4~{\rm GHz}}$ ratio. We utilise the spectral template of the archetypical starburst galaxy of M82 from the models of Efstathiou, Rowan-Robinson $\&$ Siebenmorgen (2000) for the spectral energy distribution of the star-forming population. The radio and far-infrared fluxes are correlated due to the presence of hot OB stars in giant molecular clouds that heat the surrounding dust producing the infrared emission. These stars subsequently end their lives as supernovae with the radio emission powered by the synchrotron emission from their remnants. The radio spectrum is characterised by a power law of ($S_{\nu}\propto \nu ^{-\alpha}$, $\alpha$=0.8).

\begin{figure}
 \centering
 \includegraphics[width=1.0\linewidth]{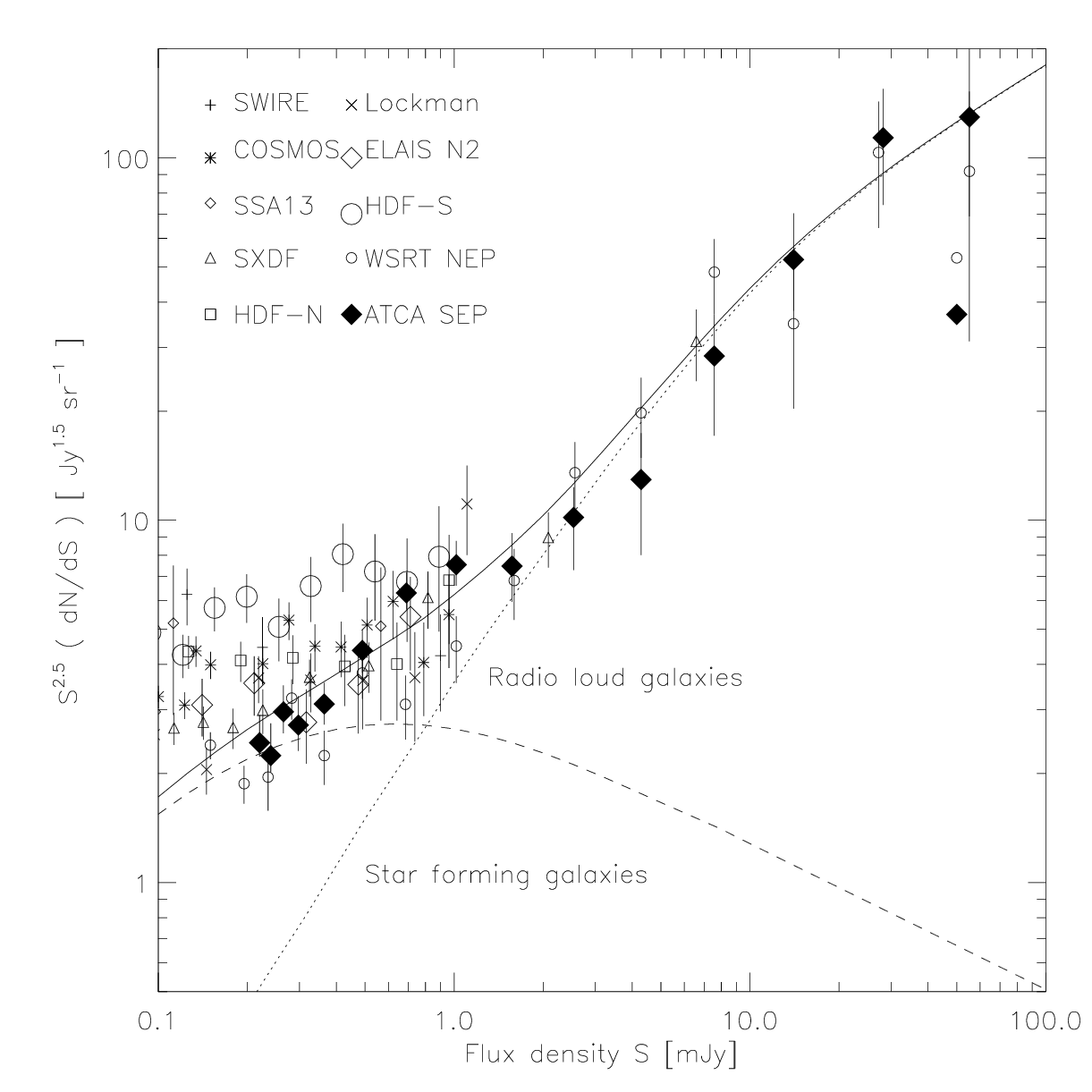}\\
 \caption{A compilation of the differential source counts of a number of deep 20 cm radio surveys taken from: SWIRE (Owen $\&$ Morrison 2008); COSMOS (Bondi \al 2008); SSA13 (Fomalont \al 2006); SXDF (Simpson \al 2006); HDF-N, LOCKMAN and ELAIS N2 (Biggs $\&$ Ivison 2006), and the HDF-S (Huynh \al 2005). The solid curve is the best fit to the present data taken as described in Figure \ref{ATCA_diff_counts}. There are however differences in the instrumental and systematic corrections that have been made for the different survey results shown here (see detailed discussion by Prandoni \al 2000 a,b), that make quantitative comparison at the faintest flux levels somewhat uncertain.}
 \label{Fig:plot_of_plots}
\end{figure}

Pure luminosity evolution for the star-forming population is assumed with a best fit power law $\propto (1+z)^{3.2}$. This infrared representation of the star-forming population was preferred over using the radio luminosity function directly, since it creates a phenomenological link between the radio emission and the infrared which is responsible for the bulk of the emission in the star-forming population. The observed number counts at fainter fluxes ($<$1mJy) vary widely from survey to survey resulting in a distribution of the best fitting evolution parameterisation. Huynh \al (2005) used the radio luminosity function of Condon \al (2002) and derived a best fitting evolution parameterisation $\propto (1+z)^{2.7}$, slightly lower than the work presented here. Hopkins (2004) and Hopkins \al (1998) used radio and infrared luminosity functions respectively obtaining evolution in the sub-mJy population $\propto (1+z)^{2.7}$ and $\propto (1+z)^{3.3}$ respectively. Comparing our observations and assumed evolution with the results of our survey in the {\it AKARI} deep field at the North Ecliptic Pole (Paper 1) we find that our derived evolution for the AGN and star forming components ($(1+z)^{3.0}$, $(1+z)^{3.2}$ respectively) are consistent with the values arrived at for the survey at the North Ecliptic Pole ($(1+z)^{3.0}$ for both components). Both of our surveys at both ecliptic poles (each covering areas of $\ga$1 deg$^{2}$, similar to the VLA-COSMOS survey of Bondi \al (2008) and larger than the other surveys depicted in Figure \ref{Fig:plot_of_plots}) result in number counts  at the lower end of the emerging picture on excess sub-mJy radio counts, as shown in Figure \ref{Fig:plot_of_plots}.

\section{Cross-matches with deep field catalogues at other wavelengths}
To compare the ATCA radio catalogue with the AKARI FIS 90~$\mu$m ADF-S catalogue (Shirahata \al ${\it in~preparation}$), we have cross-matched AKARI sources and radio components to search for positional coincidences $\le$6\arcsec ~based on the AKARI positional uncertainty as defined in Verdugo \al (2007) and the radio components reported in this paper. In the case of the possible double or complex radio sources (see Figures \ref{nearestneighbour} and \ref{ATCAresolvedsources}) we have also searched for candidate identification along a line joining the presumed associated radio components. The AKARI FIS catalogue covers the entire 12 deg$^{2}$ of the ADF-S and contains  391, 2282, 315 and 216 sources at 65~$\mu$m, 90~$\mu$m, 140~$\mu$m, 160~$\mu$m above  5$\sigma$ detection sensitivities of 28.47 mJy, 12.81 mJy, 121.03 mJy and 372.68 mJy respectively.  From this cross matching we recovered 35 sources in common to both catalogues, twenty-five of which are also reported in the ${\it Spitzer}$ 70~$\mu$m catalogue (Clements \al 2011). We list the ATCA-AKARI cross-matched sources in Table \ref{Table:crossmatches}, along with R-band detections from our CTIO MOSAIC-II survey (or in a few cases that are marked with a {\it dagger} symbol from DENIS R-band fluxes), and redshifts from the AAT/AAOmega redshift survey. The ${\it Herschel}$ data are described in the Figure caption of Table \ref{Table:crossmatches}. The 41 cross matched sources, all of which lay within a 5$^{\prime\prime}$ error circle, the mean positional agreement was 1.93$\pm$1.25$^{\prime\prime}$, showing very good agreement of the coordinate systems.


\begin{table*}
\caption{AKARI, ${\it Spitzer}$ and ${\it Herschel}$ associations with radio sources in the ATCA-ADFS survey. The ${\it Herschel}$ (referenced as HSO in this Table) HerMES fluxes were extracted from the ${\it Herschel}$ HerMES Public data release available from the LEDAM server at http://hedam.oamp.fr/HerMES/release.php, where we have used the band-merged StarFinder catalogues with the xID multi-band (250, 350 and 500 micron) fluxes measured at the positions of the StarFinder 250 micron
sources.  }
\begin{tabular}{c c c c c c c c c c c c c c c c}
\hline
\multicolumn{1}{c}{No} & \multicolumn{1}{c}{S$_{20cm}$} & \multicolumn{1}{c}{CTIO} &\multicolumn{1}{c}{J} & \multicolumn{1}{c}{H} & \multicolumn{1}{c}{K} & \multicolumn{1}{c}{${\it Spitzer}$} & \multicolumn{1}{c}{${\it Spitzer}$} & \multicolumn{1}{c}{AKARI} & \multicolumn{1}{c}{AKARI} & \multicolumn{1}{c}{AKARI} & \multicolumn{1}{c}{HSO} &\multicolumn{1}{c}{HSO} &\multicolumn{1}{c}{HSO} &  \multicolumn{1}{c}{Redshift}  \\ 
\multicolumn{1}{c}{} & \multicolumn{1}{c}{mJy} & \multicolumn{1}{c}{R mag} & \multicolumn{1}{c}{mag} & \multicolumn{1}{c}{mag} & \multicolumn{1}{c}{mag} & \multicolumn{1}{c}{24~$\mu$m} & \multicolumn{1}{c}{70~$\mu$m} & \multicolumn{1}{c}{65~$\mu$m} & \multicolumn{1}{c}{90~$\mu$m} & \multicolumn{1}{c}{140~$\mu$m} & \multicolumn{1}{c}{250~$\mu$m} & \multicolumn{1}{c}{350~$\mu$m} & \multicolumn{1}{c}{500~$\mu$m} &  \multicolumn{1}{c}{${\it z}$} \\ 
\multicolumn{1}{c}{} & \multicolumn{1}{c}{} & \multicolumn{1}{c}{} & \multicolumn{1}{c}{} & \multicolumn{1}{c}{} & \multicolumn{1}{c}{} & \multicolumn{1}{c}{mJy} & \multicolumn{1}{c}{mJy} & \multicolumn{1}{c}{mJy} & \multicolumn{1}{c}{mJy} & \multicolumn{1}{c}{mJy} & \multicolumn{1}{c}{mJy} & \multicolumn{1}{c}{mJy} &  \multicolumn{1}{c}{mJy}  \\ 
\hline
18 & 2.291 & 14.1${\bf ^{\dagger}}$ & 13.7 & 13.2 & 12.0 & 12.8 & 118.1 & 84.0 & 126.8 & - & - &- & - & 0.044\\
55 & 0.317 & 15.9${\bf ^{\dagger}}$ & 14.9 & 14.0 & 13.9 & 5.4 & 43.3 & 41.6 & 58.2 & - & 91.9 &39.6 & 8.4 & 0.092\\
63 & 0.335 &17.9${\bf ^{\dagger}}$ & - & - & - & 0.7 & 35.8 & - & 53.4 & - & 115.9 &63.7 &29.5 & -\\
72 & 0.257 & - & - & - & - & - & - & - & - & - & - &- & - & 0.423\\
77 & 0.419 & 20.6 & - & - & - & 1.1 & - & - & 15.5 & - & 84.8 &52.3 & 28.8 & -\\
83 & 0.321 & 18.9 & - & - & - & - & - & - & - & - & 64.6 &42.1 & 23.6 & -\\
84 & 0.234 & 18.9 & - & - & - & - & - & - & - & - & - &- & - & 0.329\\
100 & 0.476 & 22.0 & - & - & - & 0.8 & 44.1 & - & 51.2 & - & 77.9 &39.4 & 8.6 & -\\
112 & 0.890  & - & - & - & - & 1.1 & 49.3 & - & 90.1 & - & 69.1&38.3 & 23.4 & -\\
120 & 0.203 & 21.4 & - & - & - & - & - & - & - & - & 58.4 &40.4 & 32.4 & 0.825\\
125 & 0.415 & 16.9 & 16.4 & 15.8 & 14.9 & 5.2 & 91.5 & - & 91.2 & - & 115.9 &63.6 & 29.5 & 0.164\\
132 & 0.257 & 23.6 & - & - & - & 3.0 & - & - & 37.4 & - & 104.5 &74.7 & 37.6 & -\\
138 & 0.464 & 18.6 & - & - & - & - & - & - & - & - & - &- & - & 0.472\\
143 & 0.381 & 21.5 & - & - & - & - & - & - & 19.1 & - & - &- & - & 0.732\\
146 & 0.401 & - & - & - & - & 2.1 & 38.8 & - & 45.6 & - & 85.7 &42.2 & 11.0 & 0.393\\
159 & 0.210 & 20.5 & - & - & - & - & - & - & - & - & - &- & - & 0.577\\
164 & 0.258 & 17.6 & 16.6 & 15.8 & 15.2 & 1.5 & - & - & 22.8 & - &- & - & - & -\\
168 & 0.273 & - & - & - & - & - & - & - & - & - &58.9 & 37.4 & 8.3 & -\\
170 & 0.312  & 21.3 & - & - & - & 3.1 & 69.2 & - & 77.8 & - & 86.3 &48.2 & 20.9 & -\\
184 & 0.457 & 19.2 & - & - & - & - & - & - & - & - & - &- & - & 0.327\\
187 & 8.675 &17.9${\bf ^{\dagger}}$ & 16.4 & - & - & 22.0 & 35.4 & - & 19.3 & - &- & - & - & 0.121\\
194 & 0.510 & 19.2 & - & - & - & 5.4 & 88.8 & 44.8 & 96.9 & - & 64.1 &22.2 & 1.8 & 0.290\\
195 & 0.492 & 17.9 & 16.5 & - & - & 6.0 & 49.2 & - & 48.1 & - & - &- & - & 0.237\\
203 & 0.361 & - & - & - & - & - & - & - & - & - & 64.8 &47.3 & 26.4 & -\\
206 & 0.952 & 17.5 & - & - & - & - & - & - & - & - & 75.1 &45.2 & 18.6 & 0.361\\
214 & 0.276 & 14.3 & 13.4 & - & 12.2 & 3.7 & 105.5 & - & 115.9 & - &145.9 & 58.4 & 13.7 & -\\
225 & 0.177 & - & - & - & - & - & - & - & - & - & - &- & - & 0.181\\
235 & 0.942 & 23.1 & - & - & - & 2.5 & - & - & 24.7 & - & 86.2 &59.2 & 31.5 & -\\
252 & 0.350 & 22.3 & - & - & - & 6.7 & 36.7 & - & 36.4 & - & - &- & - & 0.154\\
255 & 0.295 & 22.3 & - & - & - & - & - & - & - & - & 62.9 &39.7 & 22.4 & -\\
266 & 0.929 & - & - & - & - & - & - & - & - & - & 55.5 &45.1 & 31.9 & -\\
277 & 0.798 & 15.3 & 14.5 & - & 13.3 & 2.6 & 80.0 & - & 103.2 & 245.0 &158.7 & 68.6 & 20.4 & -\\
283 & 0.300 & - & - & - & - & 0.2 & - & - & 15.3 & - & - &- & - & -\\
284 & 0.594 & - & - & - & - & 3.8 & 67.2 & - & 65.9 & - & 75.3 &39.8 & 24.6 & -\\
290 & 1.182 & - & - & - & - & - & - & - & - & - & 59.2 &45.3 & 22.1 & -\\
297 & 0.270 & 17.8 & 16.6 & 15.8 & 15.5 & 3.3 & 45.8 & 49.4 & 63.0 & - &- & - & - & 0.069\\
301 & 0.154 & 18.8 & - & - & - & - & - & - & 40.5 & - & - &- & - & 0.223\\
302 & 0.273 & 21.2 & - & - & - & - & - & - & 13.6 & - & - &- & - & 0.760\\
309 & 0.215 & - & - & - & - & 1.3 & - & - & 25.7 & - & - &- & - & -\\
330 & 0.263 & - & - & - & - & 7.7 & 53.3 & - & 40.2 & - & - &- & - & 0.181\\
333 & 0.184 & - & - & - & - & - & - & - & 27.2 & - & 64.3 &32.7 & 11.1 & 0.108\\
336 & 0.301 & - & - & - & - & - & - & - & - & - & 191.2 &83.9 & 10.1 & -\\
337 & 0.334 & 17.1 & 13.6 & 12.8 & 12.3 & - & 112.1 & 91.9 & 161.5 & 439.0 &186.4 & 82.8 & 52.8 & 0.0463\\
341 & 0.166 & - & 14.7 & 13.8 & 13.6 & 5.0 & 69.6 & - & 101.3 & - &87.8 & 39.0 & 16.2 & -\\
346 & 0.269 & - & - & - & - & 0.6 & - & - & 29.2 & - & - &- & - & -\\
350 & 0.282 & - & - & - & - & 1.5 & 37.9 & - & 48.9 & - & - &- & - & -\\
358 & 0.798 & - & - & - & - & - & - & - & 143.0 & - & 110.7 &48.9 & 12.6 & 0.1140\\
359 & 0.397 & - & - & - & - & - & - & - & - & - & 59.6 &46.7 & 28.0 & -\\

\hline \hline
\end{tabular}
\label{Table:crossmatches}
\end{table*}



\begin{table*}
\caption{Table continues from above: AKARI, ${\it Spitzer}$ and ${\it Herschel}$ associations with radio sources in the ATCA-ADFS survey. The ${\it Herschel}$ fluxes were extracted from the ${\it Herschel}$ HerMES Public data release available from the LEDAM server at http://hedam.oamp.fr/HerMES/release.php, where we have used the band-merged StarFinder catalogues with xID multi-band (250~$\mu$m, 350~$\mu$m and 500~$\mu$m) fluxes measured at the positions of the StarFinder 250~$\mu$m
sources.  }
\begin{tabular}{c c c c c c c c c c c c c c c c}
\hline
\multicolumn{1}{c}{No} & \multicolumn{1}{c}{S$_{20cm}$} & \multicolumn{1}{c}{CTIO} &\multicolumn{1}{c}{J} & \multicolumn{1}{c}{H} & \multicolumn{1}{c}{K} & \multicolumn{1}{c}{${\it Spitzer}$} & \multicolumn{1}{c}{${\it Spitzer}$} & \multicolumn{1}{c}{AKARI} & \multicolumn{1}{c}{AKARI} & \multicolumn{1}{c}{AKARI} & \multicolumn{1}{c}{HSO} &\multicolumn{1}{c}{HSO} &\multicolumn{1}{c}{HSO} &  \multicolumn{1}{c}{Redshift}  \\ 
\multicolumn{1}{c}{} & \multicolumn{1}{c}{mJy} & \multicolumn{1}{c}{R mag} & \multicolumn{1}{c}{mag} & \multicolumn{1}{c}{mag} & \multicolumn{1}{c}{mag} & \multicolumn{1}{c}{24~$\mu$m} & \multicolumn{1}{c}{70~$\mu$m} & \multicolumn{1}{c}{65~$\mu$m} & \multicolumn{1}{c}{90~$\mu$m} & \multicolumn{1}{c}{140~$\mu$m} & \multicolumn{1}{c}{250~$\mu$m} & \multicolumn{1}{c}{350~$\mu$m} & \multicolumn{1}{c}{500~$\mu$m} &  \multicolumn{1}{c}{${\it z}$} \\ 
\multicolumn{1}{c}{} & \multicolumn{1}{c}{} & \multicolumn{1}{c}{} & \multicolumn{1}{c}{} & \multicolumn{1}{c}{} & \multicolumn{1}{c}{} & \multicolumn{1}{c}{mJy} & \multicolumn{1}{c}{mJy} & \multicolumn{1}{c}{mJy} & \multicolumn{1}{c}{mJy} & \multicolumn{1}{c}{mJy} & \multicolumn{1}{c}{mJy} & \multicolumn{1}{c}{mJy} &  \multicolumn{1}{c}{mJy}  \\ 
\hline

410 & 0.301 & 16.3${\bf ^{\dagger}}$ & 16.5 & 16.0 & 15.3 & 2.2 & 45.2 & - & 46.8 & - & 88.7 &35.4 & 25.1 & 0.413\\
419 & 0.409 & 17.7${\bf ^{\dagger}}$ & - & - & - & 3.6 & 66.7 & 48.9 & 71.7 & - & 72.7 &34.1 & - & 0.346\\
421 & 0.617 & 16.5${\bf ^{\dagger}}$ & 15.9 & 15.3 & 14.5 & 5.0 & 79.3 & 62.1 & 92.1 & - & 64.5 &32.3 & 12.3 & 0.145\\
426 & 0.370 & - & - & - & - & - & - & - & 30.1 & - & 65.3 &33.2 & 16.9 & 0.159\\
452 & 0.578 & 10.3${\bf ^{\dagger}}$ & 12.4 & 11.6 & 11.4 & - & - & 20.8 & 355.9 & 618.9 &101.9 & 78.8 & - & 0.0393\\
482 & 3.386 & - & - & - & - & 0.7 & - & - & 39.1 & - & 55.9 &62.5 & 54.5 & -\\
487 & 5.249 & 17.2${\bf ^{\dagger}}$ & 16.8 & 15.9 & 15.1 & 8.1 & 70.5 & - & 81.9 & - & 62.4 &33.8 & 16.7 & 0.064\\
491 & 14.68 & - & - & - & - & 0.7 & - & - & 39.3 & - & 71.4 &31.2 & - & -\\
\hline \hline
\end{tabular}
\label{Table:crossmatchesb}
\end{table*}

\subsection{ Infrared cross matches (${\it AKARI}$, ${\it Spitzer}$)}

\begin{figure}
 \centering
 \includegraphics[width=1.37\linewidth]{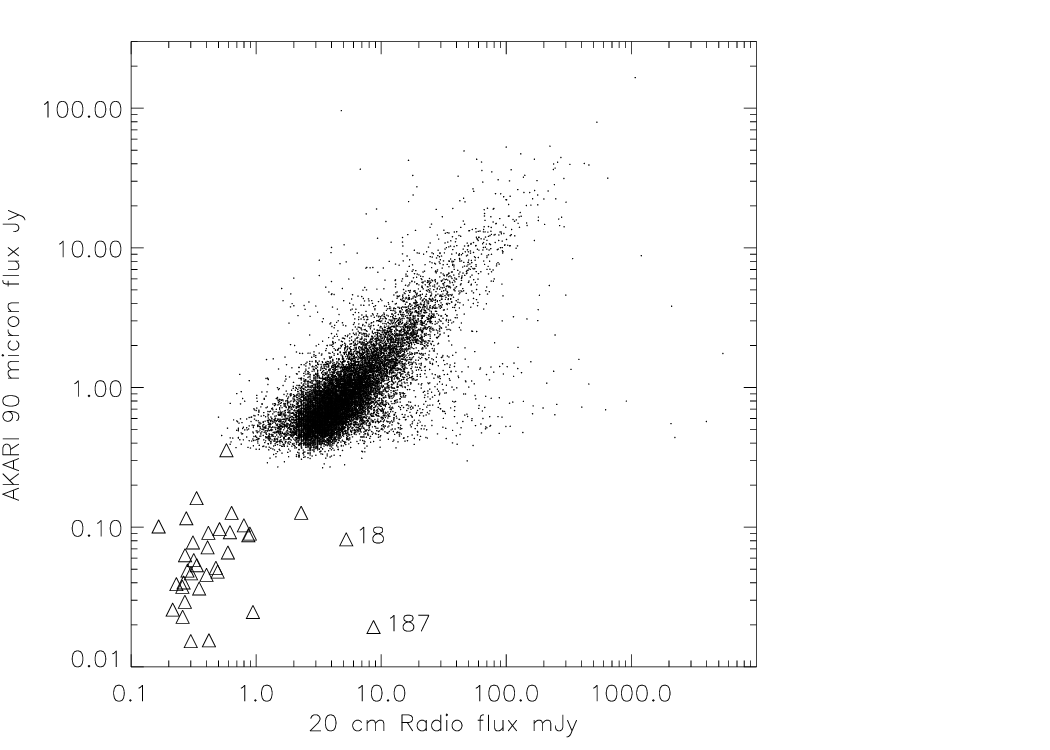}\\
 \caption{Radio flux against AKARI 90~$\mu$m flux (shown as triangles), compared to 20,000 matches between the AKARI All-Sky Bright Source Catalogue (Yamamura \al 2010), and Dixon's Master list of radio sources (shown as dots). The reason for the break between the All-Sky Survey and ADF-S FIS survey points is because of  the lower sensitivity of the All-Sky survey($\sim$ 0.4 Jy) by comparison to the deeper observations discussed here.The correlation between both FIS catalogues and Dixon's Master list of radio sources at 20 cm was made be searching for components that lay within $\pm$3\arcsec ~of each other. This conservative limit was chosen to enable secure detections, although the official AKARI FIS Bright source catalogue is 6\arcsec ~as described in Yamamura \al (2010). Radio components 18 and 187 discussed in the text are marked in the Figure.}
 \label{Fig:akari_radio_all}
\end{figure}

Figure \ref{Fig:akari_radio_all} shows the comparison between the fluxes of matched ATCA radio -- 90~$\mu$m sources detected in our survey as well as a larger sample of radio and 90~$\mu$m fluxes taken by cross-correlating with the AKARI All-Sky Survey FIS catalogue (Yamamura \al 2010, Oyabu \al 2009, 2010) with the compilation of radio sources given by Dixon (1970). Although this Figure does not apply a ${\it K}$-correction to the measured fluxes, it does show us that although many of the ATCA-ADFS sources fall on an extrapolation of sources from Dixon's list to lower fluxes, several of them may be radio loud compared to the majority (in other words lie significantly to the right of the trend line), and which therefore may have active nuclei. Of these, the two most extreme are the following. Firstly, ATCA component 18 (J04421266-5355520) at redshift 0.044 appears on the NED extragalactic database as a bright edge-on spiral galaxy with DENIS Blue and Red magnitudes of 14.6 and 14.1 respectively, and in the GALEX FUV and NUV bands with 22.39 and 21.03 mag respectively. Secondly, ATCA component 187 is a bright radio source previously detected in the SUMSS survey (SUMSS J044532-540211) with a radio flux of 1.22 mJy at 834 MHz, suggesting that it may have brightened considerably (assuming a normal spectral index), and associated with an object having DENIS Blue and Red magnitudes of 18.4 and 17.7 mag respectively and a magnitude of 21.68 in the GALEX NUV band.

\begin{figure}
 \centering
 \includegraphics[width=1.37\linewidth]{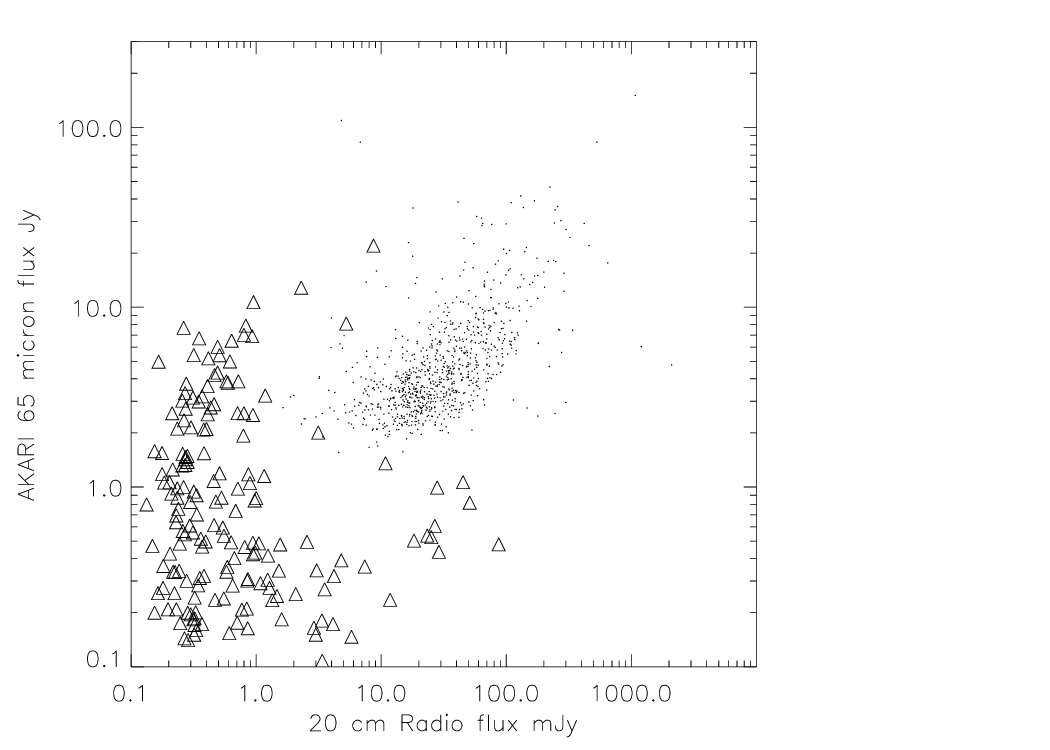}\\
 \caption{Radio components (shown as triangles) with matching ${\it Spitzer}$ 24~$\mu$m sources within one ${\it Spitzer}$ pixel. The ${\it Spitzer}$ fluxes are the point response function fitted fluxes from Scott \al (2010). The dots are matches between AKARI FIS 65~$\mu$m survey detections of bright radio sources taken from Dixon's Master radio catalogue (1970), for confirmed (i.e. quality flag 3) AKARI sources lying more than 10 degrees from the Galactic Plane as described in the caption of Figure \ref{Fig:akari_radio_all}.}. 
 \label{Fig:akari_spitzer}
\end{figure}

The radio identifications in Table \ref{sourcecatalogueshort} were cross-correlated with the ${\it Spitzer}$ 24~$\mu$m and 70~$\mu$m catalogues (Scott \al 2010), finding 173 and 31 matches at 24~$\mu$m and 70~$\mu$m respectively, using the ${\it Spitzer}$ single pixel size (2.45\arcsec and 4.0\arcsec at 24~$\mu$m and 70~$\mu$m respectively) as the search radius. The results of the 24~$\mu$m cross-matches are shown in Figure \ref{Fig:akari_spitzer}, and the large scatter of the plots highlights the difficulties of using the 24~$\mu$m fluxes as indicators of the radio flux.

This plot resembles that of Norris \al (2006) showing a wide dispersion. To check for chance associations, the radio coordinates were incremented by 60\arcsec ~in both RA and Dec, and this new list of positions was cross-correlated with the ${\it Spitzer}$ data to simulate what should be blank fields, resulting in 5 matches. Assuming that these are chance associations, the majority of the matched components ($\ge$ 97$\%$) are likely to be real associations. The brightest ${\it Spitzer}$ source shown in Figure \ref{Fig:akari_spitzer} is ATCA component 187, which is associated with an R = 17.7 magnitude galaxy, and has a redshift of 0.121 (see Table \ref{Table:crossmatches}). The ATCA components with ${\it Spitzer}$ detections which have flux densities $\ge$ 10 mJy are 11, 12, 155, 160, 236, 446, 448, 458, 530.

\subsection{Radio luminosity}
\begin{figure}
 \centering
 \includegraphics[width=1.34\linewidth]{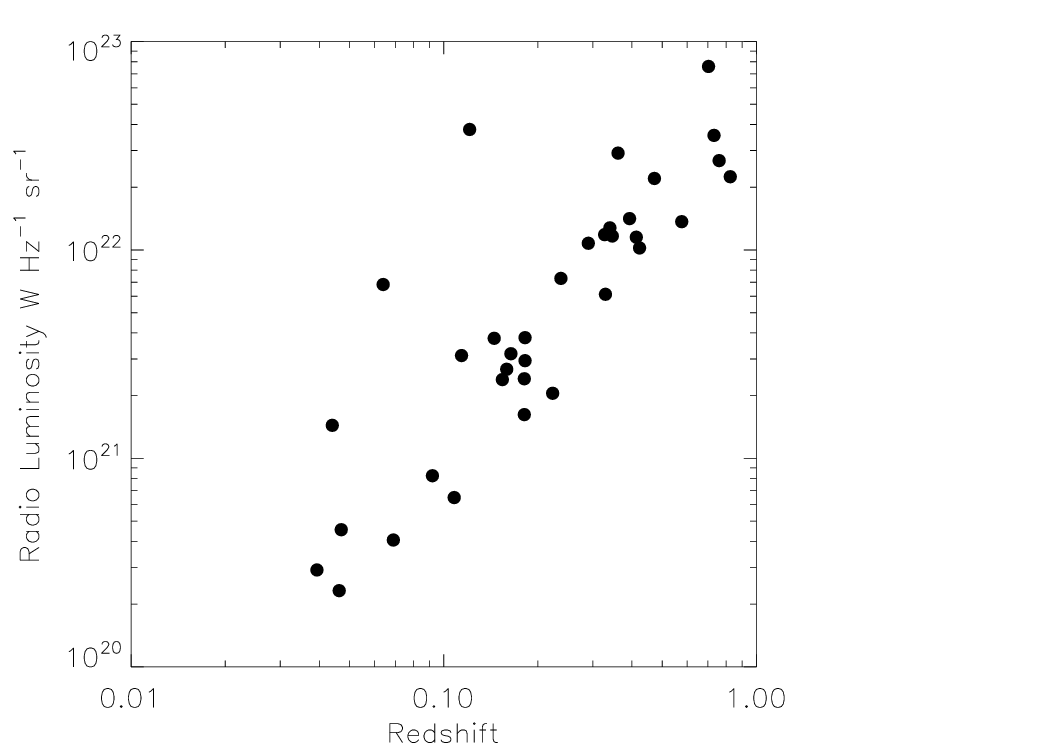}\\
 \caption{Radio luminosity as a function of the radio sources with measured spectroscopic redshifts listed in Table \ref{Table:crossmatches}.}
 \label{Fig:radio_luminosity}
\end{figure}

The radio luminosity of the sources listed in Table \ref{Table:crossmatches} was calculated, assuming a cosmology of ${\it H}$$_0$ = 70 km s$^{-1}$ Mpc$^{-1}$, with matter and cosmological constant density parameters of $\Omega$$_M$ = 0.3, $\Omega_{\Lambda}$ = 0.7. The redshifts were measured using AAOmega, the fibre-fed optical spectrograph at the Anglo Australian Observatory as described by Sedgwick \al (2011), and the resultant plot of the radio luminosity against redshift is shown in Figure \ref{Fig:radio_luminosity}, where we assume a mean radio spectral index of $\alpha$ = -0.7 (where ${\it S}$$ \propto \nu^{\alpha}$) and apply the usual form of the ${\it k}$-correction $\kappa$(${\it z}$) = $(1 + ${\it z}$)$$^{-(1 + \alpha)}$ at redshift ${\it z}$.

From studies of the local 1.4 GHz luminosity function, Sadler \al (2002) and Mauch $\&$ Sadler (2007) have shown that the low luminosity population with radio luminosity $\sim$ 10$^{23}$ W Hz$^{-1}$ will mostly be luminous star-forming galaxies rather than radio-loud AGN (Eales \al 2009, Jarvis \al 2010, Hardcastle \al 2010).  Although most of the cross-matched ATCA/AKARI sources shown in Fig. \ref{Fig:akari_radio_all} fall on the trend shown from the wider sample of cross matches between AKARI and Dixon's catalogue, ATCA components 18 and 187 appear to be radio-loud, although at the lower end of the luminosities reported from the local luminosity function for this class. However, since most of the other cross-matches fit the trend-line, we conclude that the ATCA/AKARI cross identifications primarily trace the star-forming galaxy population.

\subsection{Infrared Colours}

\begin{figure}
 \centering
 \includegraphics[width=1.0,width=1.0\linewidth]{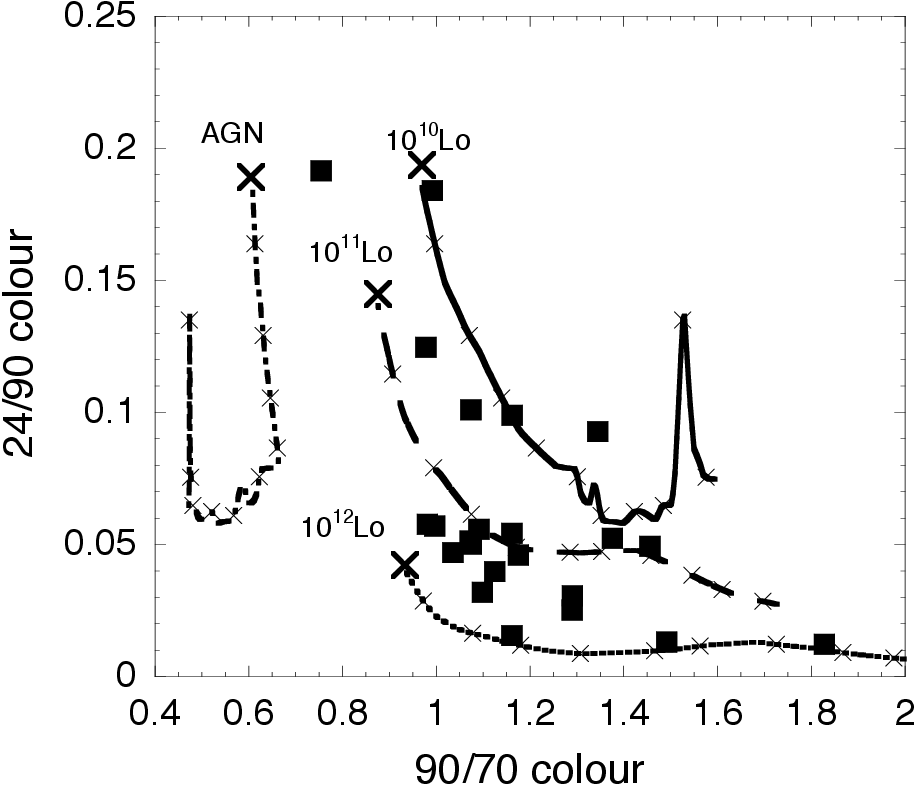}\\
 \caption{Infrared colours of the sources from our ATCA survey with cross matches in both the AKARI 90~$\mu$m and ${\it Spitzer}$ 24 \& 70~$\mu$m bands from Table \ref{Table:crossmatches}. The black squares represent the catalogue sources, with the spectral tracks of an ensemble of star-forming galaxies with increasing far-infrared luminosity from L$_{\rm IR}>10^{10}L_{\sun}$, $10^{11}L_{\sun}$, $10^{12}L_{\sun}$, and a spectral track from an AGN also plotted. The large crosses show the zero redshift points with the further smaller crosses corresponding to steps in redshift of 0.1.}
 \label{Fig:IRcolours}
\end{figure}

To further investigate the nature of the ATCA/AKARI population we compare the infrared colours of our components from our ATCA survey with cross matches in both the AKARI 90~$\mu$m and ${\it Spitzer}$ 24~$\mu$m \& 70~$\mu$m bands. In  Figure \ref{Fig:IRcolours} we plot the 24~$\mu$m~/90~$\mu$m -- 90~$\mu$m~/70~$\mu$m colour distribution of our sources. The models were derived from the SEDs smoothed by the filter bands, and are overlaid onto the spectral tracks of an ensemble of star-forming galaxies from the models of  Efstathiou \al (2000) with increasing far-infrared luminosity from L$_{\rm IR}>10^{10}L_{\sun}$, $10^{11}L_{\sun}$, $10^{12}L_{\sun}$, together with the spectral track of an AGN torus from the models of Efstathiou \& Rowan-Robinson (1995). From  Figure \ref{Fig:IRcolours} the infrared colours of the  ATCA/AKARI population are consistent with those of star-forming galaxies, although there are selection effects (the requirement to have a 90 $\mu$m cross-match) which may bias this, and which would need to be tested with more sensitive infrared observations.

As a further check, the line ratios of [OIII]/H-$\beta$ (lines at wavelengths of 486.1, 495.8, 500.7nm) and [OIII]/[OII] (OII doublet at 372.7 nm) were checked from the AAOmega spectra for the sources with redshifts $\textgreater$0.1, with the result that only one component (ATCA 302) shows ratios that are close to typical AGN values (Sedgwick \al ${\it in~ preparation}$). Therefore the radio luminosities (Figure \ref{Fig:radio_luminosity}), infrared colour-colour plots (Figure \ref{Fig:IRcolours}) and AAOmega spectra all show a consistent picture, suggesting that the ATCA/AKARI cross-identifications predominantly trace a star-bursting population.

\subsection{Optical identifications}

The positions of the components listed in Table \ref{Table:crossmatches} were compared with those in the CTIO MOSAIC-II survey (see Table \ref{ancillary}), taking a maximum search radius of 1\arcsec~ (based on the offsets to bright radio sources described in Section~\ref{bright_galaxy_offsets}. The number of galaxies as a function of R-magnitude was calculated from the CTIO MOSAIC-II survey, which covers an area of size 1.84 $\times$ 0.64 degrees centred at RA (J2000) = 4$^h$ 43$^m$ 32.8$^s$, Declination (J2000) =  -53\degree 34\arcmin 51\arcsec. Based on our choice of a radio error box of 1\arcsec~ search radius, the chance possibility that a 23rd magnitude galaxy (the most numerous in the above plot) should randomly coincide with a radio component is 0.6$\%$. Making an additional correction for the fact that some of the galaxies are extended or saturated, the chance association of a galaxy with a radio component is still $\le$ 1$\%$.

\begin{figure*}
 \centering
 \includegraphics[width=0.94\linewidth,angle=180]{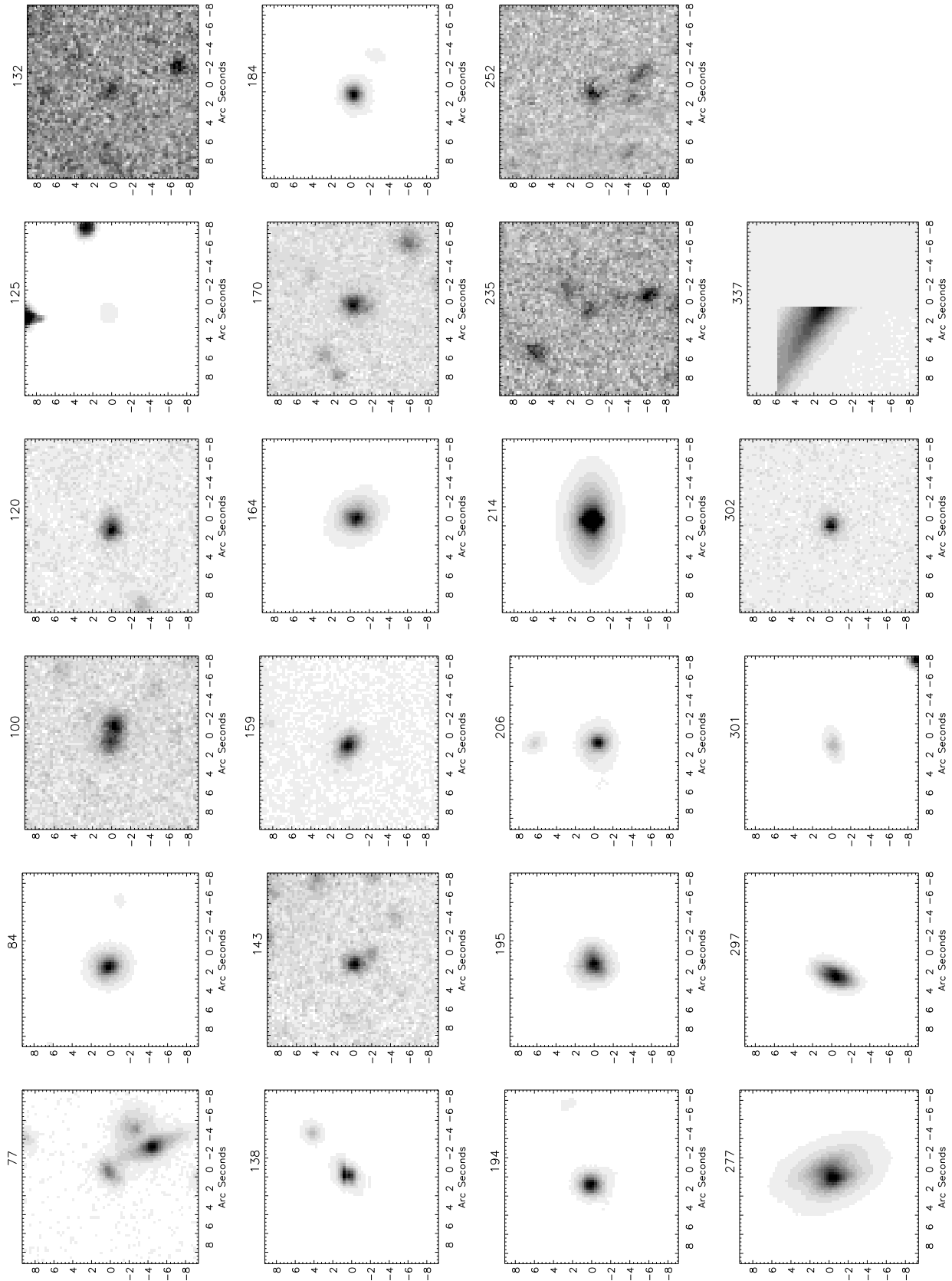}\\
 \caption{Optically identified radio components from the CTIO MOSAIC-II images. The number on top of each image is the running number of the radio component listed in Table \ref{Table:crossmatches}. The scaling of each optical image has been adjusted to show the optical galaxy.}
 \label{Fig:optical_overlays}
\end{figure*}

Postage stamp cutouts for 18$^{\prime\prime}$$\times$18$^{\prime\prime}$ regions around the sources with CTIO MOSAIC-II matches in Table \ref{Table:crossmatches} are shown in Figure \ref{Fig:optical_overlays}, where the radio component is located at the centre of each box.

\begin{figure}
 \centering
 \includegraphics[width=0.94\linewidth]{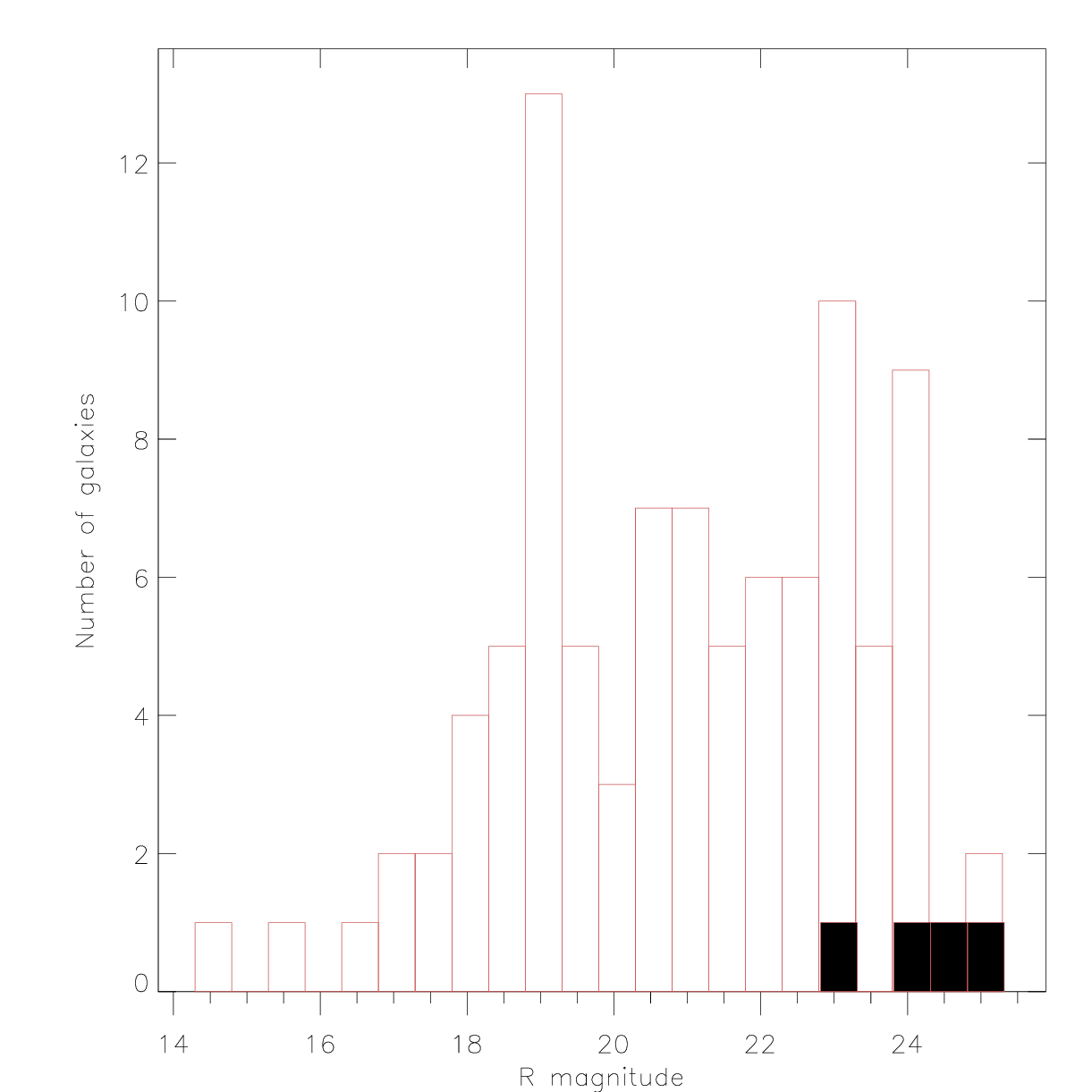}\\
 \caption{Cross correlation between the ATCA radio components and CTIO MOSAIC-II R-band galaxies within a 1\arcsec ~search radius. The equivalent counts after arbitrarily shifting the radio coordinates by +60\arcsec ~in both RA and Dec are shown as the solid filled bars at the bottom right of the Figure, as an indication of the likely false identification rate, which is clearly in line with that expected from the observed density of galaxies in the CTIO MOSAIC-II images.}
 \label{Fig:radio_ctio_all}
\end{figure}

The full radio catalogue (Table \ref{sourcecatalogueshort}) was then cross-correlated against the CTIO MOSAIC-II survey, resulting in 95 matches within a search radius of 1\arcsec. To test the false identification rate, arbitrary 60\arcsec ~offsets were again added to both the RA and Dec coordinates of the radio components, and the cross-match was repeated, resulting in only two galaxies as probably false identifications, which is roughly consistent with out estimate of the likely false detection rate discussed previously. We can therefore be confident to a high degree of the efficacy of our cross identifications. These are shown, along with the (probably false) detections arbitrarily shifting the radio pointing positions, in Figure \ref{Fig:radio_ctio_all}.

The distribution of associated galaxies with magnitude is similar to that found in the CDFS field by Simpson \al (2006) and Mainieri \al (2008), with the majority of the number of detections rising from an R-magnitude $\sim$ 17.

\subsection{Sub-millimetre cross matches}

The ATCA radio catalogue was searched for matches with the ASTE/AzTEC 1.1 mm deep survey (Hatsukade \al 2011) which contains the locations of 198 potential sub-millimetre galaxies over a $\sim$0.25 degree$^2$ area. We find one credible match that is consistent with the positional errors, lying 5.5\arcsec ~from ATCA component 120 with AzTEC J044435.35-534346.6. This has a de-boosted 1.1 mm flux of 2.8$\pm$0.5 mJy; a 20 cm radio flux 0.203 mJy; an R-magnitude of 21.4 from our CTIO imaging survey; and is at the highest redshift of 0.825 amongst the ATCA/AKARI detections detected in the AAT AAOmega redshift survey (Sedgwick \al 2011). 

We have also cross-correlated the ATCA catalogue with the BLAST South Ecliptic Pole catalogue (Valiante \al 2010) finding five cross matched associations within 10\arcsec ~of a radio position (i.e. searching to one third of the BLAST beam width). These sources, and their 250~$\mu$m fluxes are ATCA component numbers 112, 125, 168, 316 and 409 with 250~$\mu$m fluxes of 205 mJy, 119 mJy, 177 mJy, 467 mJy and 130 mJy respectively. The first two of these are also listed in Table \ref{Table:crossmatches} as having AKARI cross-matches.  We have also cross-correlated the ATCA radio sources with the ${\it Herschel}$-HerMES Public Data release catalogue, finding 41 cross-matches, the majority confirming our AKARI detections.

\section{Conclusions}
\begin{enumerate}

\item
A deep radio survey has been made of a $\sim$ 1.1 degree$^2$ area around the ATCA-ADF-S field using the ATCA telescope at 20 cm wavelength, and $\sim$ 2.5 degree$^2$ to lower sensitivity. The best sensitivity of the survey was 21 $\mu$Jy beam$^{-1}$, achieved with a synthesised beam of 6.2\arcsec ~$\times$ ~4.9$^{\prime\prime}$. The analysis methodology was carefully chosen to mitigate the various effects that can affect the efficacy of radio synthesis array observations, resulting in a final catalogue of 530 radio components, with the faintest integrated fluxes at about the 100 $\mu$Jy level. The present catalogue of radio components will form the basis of a further paper reporting cross correlation against extant AKARI and deep optical imaging. Our derived sub-mJy number counts are consistent with, but lie  at the lower end of the emerging picture for the excess in the radio counts below 1 mJy. Fitting an evolving galaxy model to our derived counts, we find a consistent picture of radio-loud dominated sources at bright fluxes and an emerging population of star-forming galaxies at radio flux levels $<$1 mJy.
\item
Cross-correlating these with far-infrared sources from AKARI, archival optical photometry, ${\it Spitzer}$ and BLAST data, we find 51 components lying within 1\arcsec ~of a radio position in at least one further catalogue. From optical identifications of a small segment of the radio image, we find 95 cross matches, with most galaxies having R-magnitudes in the range 18-24 magnitudes, similar to that found in other optical deep field identifications. The redshifts of these vary between the local universe and redshifts of up to 0.825. Associating with the ${\it Spitzer}$ catalogue, we find 173 matches at 24~$\mu$m,  within one ${\it Spitzer}$ pixel, of which a small sample are clearly radio loud compared to the bulk of the galaxies.
\item
The radio luminosity plot suggests that the majority of the radio sources with 90~$\mu$m counterparts are luminous star forming galaxies. This conclusion is supported by a comparison of the infrared colours of our matched sources which are well described by the colours expected from star-forming galaxies.
\item
There is one cross match with an ASTE source, and five cross matches with BLAST submillimetre galaxies from the radio sources detected in the present this survey, two of which are also detected also detected at by AKARI at 90~$\mu$m, and 41 detections with ${\it Herschel}$, of which 12 had not previously been identified by AKARI.

\end{enumerate}

\section{Acknowledgements}

This work is based on observations with AKARI, a ${\it JAXA}$ project with the participation of ESA. We also express our thanks to The Australia Telescope Compact Array for the substantial allocation of observing time; to the staff of the Narrabri Observatory for technical support; and the UK Science and Technology Facilities Council, ${\it STFC}$ for support. The UK-Japan AKARI Consortium has also received funding awards from the Sasakawa Foundation, The British Council, and the DAIWA Foundation, which facilitated travel and exchange activities, for which we are very grateful. This work was supported by KAKENHI (19540250 and 21111004).

\label{lastpage}

\onecolumn
\begin{table*}
\vspace{0pt}
\caption{The complete source catalogue (the full version is available as Supplementary Material in the on line version of this article). The source parameters listed in the catalog are: (1) a short form running number (components that are believed to be parts of multi-component sources are listed with a ${\bf ^{\dagger}}$ sign next to the running number (for example 47${\bf ^{\dagger}}$), (2) the source name, referred to in this paper as ATCA-ADFS followed by the RA/Dec encoding (e.g. ATCA-ADFS J045243-533127), (3,4) the source Right Ascension and Declination (J2000) referenced from the self-calibrated reference frame, (5,6) the RA and Dec errors in arc seconds, (7,8) the peak flux density, S$_{\rm peak}$, and its associated rms error, (9,10) the integrated flux densities, S$_{\rm total}$ and their  associated errors, (11, 12, 13) the major and minor axes of the fitted Gaussian source profile and orientation (major and minor axes full width at half maximum, and position angle measured east of north.}
\begin{scriptsize}
\fontsize{8}{10}\selectfont
\begin{tabular}{l l r r r r r r r r r r r}
\hline
\multicolumn{1}{l}{No} & \multicolumn{1}{c}{Source name} & \multicolumn{1}{c}{RA} & \multicolumn{1}{c}{DEC} & \multicolumn{1}{c}{$\Delta$RA} & \multicolumn{1}{c}{$\Delta$DEC} & \multicolumn{1}{c}{S$_{\rm peak}$} & \multicolumn{1}{c}{$\Delta$S$_{\rm peak}$} & \multicolumn{1}{c}{S$_{\rm total}$} & \multicolumn{1}{c}{$\Delta$S$_{\rm total}$} & \multicolumn{1}{c}{$\theta_{maj}$} & \multicolumn{1}{c}{$\theta_{min}$} & \multicolumn{1}{c}{$PA$}\\ 
 \multicolumn{1}{l}{} &         & \multicolumn{1}{c}{h:m:s.s} & \multicolumn{1}{c}{d:m:s.s} & \multicolumn{1}{c}{${\prime\prime}$} & \multicolumn{1}{c}{${\prime\prime}$} & \multicolumn{1}{c}{mJy} & \multicolumn{1}{c}{mJy} & \multicolumn{1}{c}{mJy}  & \multicolumn{1}{c}{mJy}  & \multicolumn{1}{c}{$^{\prime\prime}$} &  \multicolumn{1}{c}{$^{\prime\prime}$} &  \multicolumn{1}{c}{$\ensuremath{^\circ}\,$}\\
  \multicolumn{1}{l}{} &         & \multicolumn{1}{c}{} & \multicolumn{1}{c}{} & \multicolumn{1}{c}{} & \multicolumn{1}{c}{} & \multicolumn{1}{c}{beam$^{-1}$} & \multicolumn{1}{c}{beam$^{-1}$} & \multicolumn{1}{c}{}  & \multicolumn{1}{c}{}  & \multicolumn{1}{c}{} &  \multicolumn{1}{c}{} &  \multicolumn{1}{c}{}\\
\multicolumn{1}{l}{(1)} & \multicolumn{1}{c}{(2)} & \multicolumn{1}{c}{(3)} & \multicolumn{1}{c}{(4)} &  \multicolumn{1}{c}{(5)} & \multicolumn{1}{c}{(6)} & \multicolumn{1}{c}{(7)} & \multicolumn{1}{c}{(8)} & \multicolumn{1}{c}{(9)} & \multicolumn{1}{c}{(10)} & \multicolumn{1}{c}{(11)}  & \multicolumn{1}{c}{(12)} & \multicolumn{1}{c}{(13)} \\
\hline
 1 &  ATCA-ADFS J044041-534043 & 4:40:41.5 & -53:40:43.5 & 0.14 & 0.03 &   2.951 &   0.286 &   7.392 &   0.368 &   12.1 &    2.7 & -84.0\\
 2 &  ATCA-ADFS J044116-532554 & 4:41:16.4 & -53:25:54.0 & 0.00 & 0.00 &   1.498 &   0.153 &   4.427 &   0.206 &    8.7 &    6.6 & -48.1\\
 3 &  ATCA-ADFS J044116-531845 & 4:41:16.9 & -53:18:45.8 & 0.07 & 0.31 &   1.568 &   0.260 &   5.310 &   0.317 &   13.5 &    3.5 &  -9.3\\
\hline
 \end{tabular}
\end{scriptsize}
\label{sourcecatalogueshort_1}
\end{table*}

\begin{table*}
\vspace{0pt}
\caption{Continuation of the source catalogue}
\begin{scriptsize}
\fontsize{8}{10}\selectfont
\begin{tabular}{l l r r r r r r r r r r r}
\hline
\multicolumn{1}{l}{No} & \multicolumn{1}{c}{Source name} & \multicolumn{1}{c}{RA} & \multicolumn{1}{c}{DEC} & \multicolumn{1}{c}{$\Delta$RA} & \multicolumn{1}{c}{$\Delta$DEC} & \multicolumn{1}{c}{S$_{\rm peak}$} & \multicolumn{1}{c}{$\Delta$S$_{\rm peak}$} & \multicolumn{1}{c}{S$_{\rm total}$} & \multicolumn{1}{c}{$\Delta$S$_{\rm total}$} & \multicolumn{1}{c}{$\theta_{maj}$} & \multicolumn{1}{c}{$\theta_{min}$} & \multicolumn{1}{c}{$PA$}\\ 
 \multicolumn{1}{l}{} &         & \multicolumn{1}{c}{h:m:s.s} & \multicolumn{1}{c}{d:m:s.s} & \multicolumn{1}{c}{${\prime\prime}$} & \multicolumn{1}{c}{${\prime\prime}$} & \multicolumn{1}{c}{mJy} & \multicolumn{1}{c}{mJy} & \multicolumn{1}{c}{mJy}  & \multicolumn{1}{c}{mJy}  & \multicolumn{1}{c}{$^{\prime\prime}$} &  \multicolumn{1}{c}{$^{\prime\prime}$} &  \multicolumn{1}{c}{$\ensuremath{^\circ}\,$}\\
  \multicolumn{1}{l}{} &         & \multicolumn{1}{c}{} & \multicolumn{1}{c}{} & \multicolumn{1}{c}{} & \multicolumn{1}{c}{} & \multicolumn{1}{c}{beam$^{-1}$} & \multicolumn{1}{c}{beam$^{-1}$} & \multicolumn{1}{c}{}  & \multicolumn{1}{c}{}  & \multicolumn{1}{c}{} &  \multicolumn{1}{c}{} &  \multicolumn{1}{c}{}\\
\multicolumn{1}{l}{(1)} & \multicolumn{1}{c}{(2)} & \multicolumn{1}{c}{(3)} & \multicolumn{1}{c}{(4)} &  \multicolumn{1}{c}{(5)} & \multicolumn{1}{c}{(6)} & \multicolumn{1}{c}{(7)} & \multicolumn{1}{c}{(8)} & \multicolumn{1}{c}{(9)} & \multicolumn{1}{c}{(10)} & \multicolumn{1}{c}{(11)}  & \multicolumn{1}{c}{(12)} & \multicolumn{1}{c}{(13)} \\
\hline

 51 &  ATCA-ADFS J044316-534148 & 4:43:16.6 & -53:41:48.4 & 0.00 & 0.00 &   2.265 &   0.038 &   3.040 &   0.059 &  &  & \\
 52 &  ATCA-ADFS J044317-524701 & 4:43:17.1 & -52:47:01.5 & 0.01 & 0.01 &   8.447 &   1.066 &   8.516 &   1.113 &  &  & \\
 53 &  ATCA-ADFS J044317-533807 & 4:43:17.9 & -53:38:07.7 & 0.00 & 0.00 &   0.641 &   0.038 &   0.851 &   0.040 &  &  & \\
\hline
 \end{tabular}
\end{scriptsize}
\label{sourcecatalogueshort_2}
\end{table*}

\begin{table*}
\vspace{0pt}
\caption{Continuation of the source catalogue}
\begin{scriptsize}
\fontsize{8}{10}\selectfont
\begin{tabular}{l l r r r r r r r r r r r}
\hline
\multicolumn{1}{l}{No} & \multicolumn{1}{c}{Source name} & \multicolumn{1}{c}{RA} & \multicolumn{1}{c}{DEC} & \multicolumn{1}{c}{$\Delta$RA} & \multicolumn{1}{c}{$\Delta$DEC} & \multicolumn{1}{c}{S$_{\rm peak}$} & \multicolumn{1}{c}{$\Delta$S$_{\rm peak}$} & \multicolumn{1}{c}{S$_{\rm total}$} & \multicolumn{1}{c}{$\Delta$S$_{\rm total}$} & \multicolumn{1}{c}{$\theta_{maj}$} & \multicolumn{1}{c}{$\theta_{min}$} & \multicolumn{1}{c}{$PA$}\\ 
 \multicolumn{1}{l}{} &         & \multicolumn{1}{c}{h:m:s.s} & \multicolumn{1}{c}{d:m:s.s} & \multicolumn{1}{c}{${\prime\prime}$} & \multicolumn{1}{c}{${\prime\prime}$} & \multicolumn{1}{c}{mJy} & \multicolumn{1}{c}{mJy} & \multicolumn{1}{c}{mJy}  & \multicolumn{1}{c}{mJy}  & \multicolumn{1}{c}{$^{\prime\prime}$} &  \multicolumn{1}{c}{$^{\prime\prime}$} &  \multicolumn{1}{c}{$\ensuremath{^\circ}\,$}\\
  \multicolumn{1}{l}{} &         & \multicolumn{1}{c}{} & \multicolumn{1}{c}{} & \multicolumn{1}{c}{} & \multicolumn{1}{c}{} & \multicolumn{1}{c}{beam$^{-1}$} & \multicolumn{1}{c}{beam$^{-1}$} & \multicolumn{1}{c}{}  & \multicolumn{1}{c}{}  & \multicolumn{1}{c}{} &  \multicolumn{1}{c}{} &  \multicolumn{1}{c}{}\\
\multicolumn{1}{l}{(1)} & \multicolumn{1}{c}{(2)} & \multicolumn{1}{c}{(3)} & \multicolumn{1}{c}{(4)} &  \multicolumn{1}{c}{(5)} & \multicolumn{1}{c}{(6)} & \multicolumn{1}{c}{(7)} & \multicolumn{1}{c}{(8)} & \multicolumn{1}{c}{(9)} & \multicolumn{1}{c}{(10)} & \multicolumn{1}{c}{(11)}  & \multicolumn{1}{c}{(12)} & \multicolumn{1}{c}{(13)} \\
\hline

 101 &  ATCA-ADFS J044419-535132 & 4:44:19.3 & -53:51:32.7 & 0.00 & 0.01 &   0.224 &   0.032 &   0.239 &   0.032 &  &  & \\
 102${\bf ^{\dagger}}$ &  ATCA-ADFS J044419-532442 & 4:44:19.9 & -53:24:42.7 & 0.07 & 0.04 &   2.513 &   0.031 &   9.265 &   0.241 &   12.9 &    5.9 & -55.1\\
 103 &  ATCA-ADFS J044420-534032 & 4:44:20.8 & -53:40:32.8 & 0.01 & 0.01 &   0.178 &   0.030 &   0.197 &   0.030 &  &  & \\
\hline
 \end{tabular}
\end{scriptsize}
\label{sourcecatalogueshort_3}
\end{table*}

\begin{table*}
\vspace{0pt}
\caption{Continuation of the source catalogue}
\begin{scriptsize}
\fontsize{8}{10}\selectfont
\begin{tabular}{l l r r r r r r r r r r r}
\hline
\multicolumn{1}{l}{No} & \multicolumn{1}{c}{Source name} & \multicolumn{1}{c}{RA} & \multicolumn{1}{c}{DEC} & \multicolumn{1}{c}{$\Delta$RA} & \multicolumn{1}{c}{$\Delta$DEC} & \multicolumn{1}{c}{S$_{\rm peak}$} & \multicolumn{1}{c}{$\Delta$S$_{\rm peak}$} & \multicolumn{1}{c}{S$_{\rm total}$} & \multicolumn{1}{c}{$\Delta$S$_{\rm total}$} & \multicolumn{1}{c}{$\theta_{maj}$} & \multicolumn{1}{c}{$\theta_{min}$} & \multicolumn{1}{c}{$PA$}\\ 
 \multicolumn{1}{l}{} &         & \multicolumn{1}{c}{h:m:s.s} & \multicolumn{1}{c}{d:m:s.s} & \multicolumn{1}{c}{${\prime\prime}$} & \multicolumn{1}{c}{${\prime\prime}$} & \multicolumn{1}{c}{mJy} & \multicolumn{1}{c}{mJy} & \multicolumn{1}{c}{mJy}  & \multicolumn{1}{c}{mJy}  & \multicolumn{1}{c}{$^{\prime\prime}$} &  \multicolumn{1}{c}{$^{\prime\prime}$} &  \multicolumn{1}{c}{$\ensuremath{^\circ}\,$}\\
  \multicolumn{1}{l}{} &         & \multicolumn{1}{c}{} & \multicolumn{1}{c}{} & \multicolumn{1}{c}{} & \multicolumn{1}{c}{} & \multicolumn{1}{c}{beam$^{-1}$} & \multicolumn{1}{c}{beam$^{-1}$} & \multicolumn{1}{c}{}  & \multicolumn{1}{c}{}  & \multicolumn{1}{c}{} &  \multicolumn{1}{c}{} &  \multicolumn{1}{c}{}\\
\multicolumn{1}{l}{(1)} & \multicolumn{1}{c}{(2)} & \multicolumn{1}{c}{(3)} & \multicolumn{1}{c}{(4)} &  \multicolumn{1}{c}{(5)} & \multicolumn{1}{c}{(6)} & \multicolumn{1}{c}{(7)} & \multicolumn{1}{c}{(8)} & \multicolumn{1}{c}{(9)} & \multicolumn{1}{c}{(10)} & \multicolumn{1}{c}{(11)}  & \multicolumn{1}{c}{(12)} & \multicolumn{1}{c}{(13)} \\
\hline
 
 151 &  ATCA-ADFS J044504-531756 & 4:45:04.6 & -53:17:56.5 & 1.11 & 0.08 &   0.211 &   0.032 &   0.133 &   0.032 &  &  & \\
 152 &  ATCA-ADFS J044504-525431 & 4:45:04.9 & -52:54:31.9 & 0.01 & 0.02 &   0.629 &   0.100 &   1.304 &   0.103 &    7.8 &    2.8 &   0.3\\
 153 &  ATCA-ADFS J044506-531726 & 4:45:06.9 & -53:17:26.8 & 0.00 & 0.01 &   0.191 &   0.032 &   0.265 &   0.032 &    7.5 &  &  -4.6\\
\hline
 \end{tabular}
\end{scriptsize}
\label{sourcecatalogueshort_4}
\end{table*}

\begin{table*}
\vspace{0pt}
\caption{Continuation of the source catalogue}
\begin{scriptsize}
\fontsize{8}{10}\selectfont
\begin{tabular}{l l r r r r r r r r r r r}
\hline
\multicolumn{1}{l}{No} & \multicolumn{1}{c}{Source name} & \multicolumn{1}{c}{RA} & \multicolumn{1}{c}{DEC} & \multicolumn{1}{c}{$\Delta$RA} & \multicolumn{1}{c}{$\Delta$DEC} & \multicolumn{1}{c}{S$_{\rm peak}$} & \multicolumn{1}{c}{$\Delta$S$_{\rm peak}$} & \multicolumn{1}{c}{S$_{\rm total}$} & \multicolumn{1}{c}{$\Delta$S$_{\rm total}$} & \multicolumn{1}{c}{$\theta_{maj}$} & \multicolumn{1}{c}{$\theta_{min}$} & \multicolumn{1}{c}{$PA$}\\ 
 \multicolumn{1}{l}{} &         & \multicolumn{1}{c}{h:m:s.s} & \multicolumn{1}{c}{d:m:s.s} & \multicolumn{1}{c}{${\prime\prime}$} & \multicolumn{1}{c}{${\prime\prime}$} & \multicolumn{1}{c}{mJy} & \multicolumn{1}{c}{mJy} & \multicolumn{1}{c}{mJy}  & \multicolumn{1}{c}{mJy}  & \multicolumn{1}{c}{$^{\prime\prime}$} &  \multicolumn{1}{c}{$^{\prime\prime}$} &  \multicolumn{1}{c}{$\ensuremath{^\circ}\,$}\\
  \multicolumn{1}{l}{} &         & \multicolumn{1}{c}{} & \multicolumn{1}{c}{} & \multicolumn{1}{c}{} & \multicolumn{1}{c}{} & \multicolumn{1}{c}{beam$^{-1}$} & \multicolumn{1}{c}{beam$^{-1}$} & \multicolumn{1}{c}{}  & \multicolumn{1}{c}{}  & \multicolumn{1}{c}{} &  \multicolumn{1}{c}{} &  \multicolumn{1}{c}{}\\
\multicolumn{1}{l}{(1)} & \multicolumn{1}{c}{(2)} & \multicolumn{1}{c}{(3)} & \multicolumn{1}{c}{(4)} &  \multicolumn{1}{c}{(5)} & \multicolumn{1}{c}{(6)} & \multicolumn{1}{c}{(7)} & \multicolumn{1}{c}{(8)} & \multicolumn{1}{c}{(9)} & \multicolumn{1}{c}{(10)} & \multicolumn{1}{c}{(11)}  & \multicolumn{1}{c}{(12)} & \multicolumn{1}{c}{(13)} \\
\hline

 200 &  ATCA-ADFS J044553-531231 & 4:45:53.7 & -53:12:32.0 & 0.19 & 0.48 &   0.107 &   0.037 &   0.609 &   0.039 &   18.9 &    6.4 &  23.7\\
 201 &  ATCA-ADFS J044554-533522 & 4:45:54.1 & -53:35:22.4 & 0.00 & 0.00 &   2.729 &   0.031 &   3.294 &   0.038 &  &  & \\
 202 &  ATCA-ADFS J044554-534203 & 4:45:54.4 & -53:42:03.6 & 0.01 & 0.01 &   0.150 &   0.036 &   0.293 &   0.036 &    7.6 &    2.7 & -45.5\\
\hline
 \end{tabular}
\end{scriptsize}
\label{sourcecatalogueshort_5}
\end{table*}

\begin{table*}
\vspace{0pt}
\caption{Continuation of the source catalogue}
\begin{scriptsize}
\fontsize{8}{10}\selectfont
\begin{tabular}{l l r r r r r r r r r r r}
\hline
\multicolumn{1}{l}{No} & \multicolumn{1}{c}{Source name} & \multicolumn{1}{c}{RA} & \multicolumn{1}{c}{DEC} & \multicolumn{1}{c}{$\Delta$RA} & \multicolumn{1}{c}{$\Delta$DEC} & \multicolumn{1}{c}{S$_{\rm peak}$} & \multicolumn{1}{c}{$\Delta$S$_{\rm peak}$} & \multicolumn{1}{c}{S$_{\rm total}$} & \multicolumn{1}{c}{$\Delta$S$_{\rm total}$} & \multicolumn{1}{c}{$\theta_{maj}$} & \multicolumn{1}{c}{$\theta_{min}$} & \multicolumn{1}{c}{$PA$}\\ 
 \multicolumn{1}{l}{} &         & \multicolumn{1}{c}{h:m:s.s} & \multicolumn{1}{c}{d:m:s.s} & \multicolumn{1}{c}{${\prime\prime}$} & \multicolumn{1}{c}{${\prime\prime}$} & \multicolumn{1}{c}{mJy} & \multicolumn{1}{c}{mJy} & \multicolumn{1}{c}{mJy}  & \multicolumn{1}{c}{mJy}  & \multicolumn{1}{c}{$^{\prime\prime}$} &  \multicolumn{1}{c}{$^{\prime\prime}$} &  \multicolumn{1}{c}{$\ensuremath{^\circ}\,$}\\
  \multicolumn{1}{l}{} &         & \multicolumn{1}{c}{} & \multicolumn{1}{c}{} & \multicolumn{1}{c}{} & \multicolumn{1}{c}{} & \multicolumn{1}{c}{beam$^{-1}$} & \multicolumn{1}{c}{beam$^{-1}$} & \multicolumn{1}{c}{}  & \multicolumn{1}{c}{}  & \multicolumn{1}{c}{} &  \multicolumn{1}{c}{} &  \multicolumn{1}{c}{}\\
\multicolumn{1}{l}{(1)} & \multicolumn{1}{c}{(2)} & \multicolumn{1}{c}{(3)} & \multicolumn{1}{c}{(4)} &  \multicolumn{1}{c}{(5)} & \multicolumn{1}{c}{(6)} & \multicolumn{1}{c}{(7)} & \multicolumn{1}{c}{(8)} & \multicolumn{1}{c}{(9)} & \multicolumn{1}{c}{(10)} & \multicolumn{1}{c}{(11)}  & \multicolumn{1}{c}{(12)} & \multicolumn{1}{c}{(13)} \\
\hline

 251${\bf ^{\dagger}}$ &  ATCA-ADFS J044644-531752 & 4:46:44.5 & -53:17:53.0 & 0.00 & 0.01 &   1.574 &   0.042 &   3.007 &   0.067 &    8.9 &  &   0.9\\
 252 &  ATCA-ADFS J044646-531642 & 4:46:46.3 & -53:16:42.7 & 0.00 & 0.00 &   0.334 &   0.041 &   0.350 &   0.041 &  &  & \\
 253${\bf ^{\dagger}}$ &  ATCA-ADFS J044649-530014 & 4:46:49.4 & -53:00:14.7 & 0.00 & 0.00 &   7.702 &   0.064 &  14.402 &   0.115 &  &    3.7 &  -1.7\\
\hline
 \end{tabular}
\end{scriptsize}
\label{sourcecatalogueshort_6}
\end{table*}

\begin{table*}
\vspace{0pt}
\caption{Continuation of the source catalogue}
\begin{scriptsize}
\fontsize{8}{10}\selectfont
\begin{tabular}{l l r r r r r r r r r r r}
\hline
\multicolumn{1}{l}{No} & \multicolumn{1}{c}{Source name} & \multicolumn{1}{c}{RA} & \multicolumn{1}{c}{DEC} & \multicolumn{1}{c}{$\Delta$RA} & \multicolumn{1}{c}{$\Delta$DEC} & \multicolumn{1}{c}{S$_{\rm peak}$} & \multicolumn{1}{c}{$\Delta$S$_{\rm peak}$} & \multicolumn{1}{c}{S$_{\rm total}$} & \multicolumn{1}{c}{$\Delta$S$_{\rm total}$} & \multicolumn{1}{c}{$\theta_{maj}$} & \multicolumn{1}{c}{$\theta_{min}$} & \multicolumn{1}{c}{$PA$}\\ 
 \multicolumn{1}{l}{} &         & \multicolumn{1}{c}{h:m:s.s} & \multicolumn{1}{c}{d:m:s.s} & \multicolumn{1}{c}{${\prime\prime}$} & \multicolumn{1}{c}{${\prime\prime}$} & \multicolumn{1}{c}{mJy} & \multicolumn{1}{c}{mJy} & \multicolumn{1}{c}{mJy}  & \multicolumn{1}{c}{mJy}  & \multicolumn{1}{c}{$^{\prime\prime}$} &  \multicolumn{1}{c}{$^{\prime\prime}$} &  \multicolumn{1}{c}{$\ensuremath{^\circ}\,$}\\
  \multicolumn{1}{l}{} &         & \multicolumn{1}{c}{} & \multicolumn{1}{c}{} & \multicolumn{1}{c}{} & \multicolumn{1}{c}{} & \multicolumn{1}{c}{beam$^{-1}$} & \multicolumn{1}{c}{beam$^{-1}$} & \multicolumn{1}{c}{}  & \multicolumn{1}{c}{}  & \multicolumn{1}{c}{} &  \multicolumn{1}{c}{} &  \multicolumn{1}{c}{}\\
\multicolumn{1}{l}{(1)} & \multicolumn{1}{c}{(2)} & \multicolumn{1}{c}{(3)} & \multicolumn{1}{c}{(4)} &  \multicolumn{1}{c}{(5)} & \multicolumn{1}{c}{(6)} & \multicolumn{1}{c}{(7)} & \multicolumn{1}{c}{(8)} & \multicolumn{1}{c}{(9)} & \multicolumn{1}{c}{(10)} & \multicolumn{1}{c}{(11)}  & \multicolumn{1}{c}{(12)} & \multicolumn{1}{c}{(13)} \\
\hline
 
 301 &  ATCA-ADFS J044727-532626 & 4:47:27.3 & -53:26:26.6 & 0.00 & 0.00 &   0.304 &   0.032 &   0.154 &   0.032 &  &  & \\
 302 &  ATCA-ADFS J044728-532955 & 4:47:28.4 & -53:29:55.9 & 0.01 & 0.01 &   0.281 &   0.034 &   0.273 &   0.035 &  &  & \\
 303 &  ATCA-ADFS J044730-530654 & 4:47:30.7 & -53:06:54.7 & 0.10 & 0.41 &   0.076 &   0.025 &   0.230 &   0.027 &   12.6 &    2.8 &   4.8\\
\hline
 \end{tabular}
\end{scriptsize}
\label{sourcecatalogueshort_7}
\end{table*}

\begin{table*}
\vspace{0pt}
\caption{Continuation of the source catalogue}
\begin{scriptsize}
\fontsize{8}{10}\selectfont
\begin{tabular}{l l r r r r r r r r r r r}
\hline
\multicolumn{1}{l}{No} & \multicolumn{1}{c}{Source name} & \multicolumn{1}{c}{RA} & \multicolumn{1}{c}{DEC} & \multicolumn{1}{c}{$\Delta$RA} & \multicolumn{1}{c}{$\Delta$DEC} & \multicolumn{1}{c}{S$_{\rm peak}$} & \multicolumn{1}{c}{$\Delta$S$_{\rm peak}$} & \multicolumn{1}{c}{S$_{\rm total}$} & \multicolumn{1}{c}{$\Delta$S$_{\rm total}$} & \multicolumn{1}{c}{$\theta_{maj}$} & \multicolumn{1}{c}{$\theta_{min}$} & \multicolumn{1}{c}{$PA$}\\ 
 \multicolumn{1}{l}{} &         & \multicolumn{1}{c}{h:m:s.s} & \multicolumn{1}{c}{d:m:s.s} & \multicolumn{1}{c}{${\prime\prime}$} & \multicolumn{1}{c}{${\prime\prime}$} & \multicolumn{1}{c}{mJy} & \multicolumn{1}{c}{mJy} & \multicolumn{1}{c}{mJy}  & \multicolumn{1}{c}{mJy}  & \multicolumn{1}{c}{$^{\prime\prime}$} &  \multicolumn{1}{c}{$^{\prime\prime}$} &  \multicolumn{1}{c}{$\ensuremath{^\circ}\,$}\\
  \multicolumn{1}{l}{} &         & \multicolumn{1}{c}{} & \multicolumn{1}{c}{} & \multicolumn{1}{c}{} & \multicolumn{1}{c}{} & \multicolumn{1}{c}{beam$^{-1}$} & \multicolumn{1}{c}{beam$^{-1}$} & \multicolumn{1}{c}{}  & \multicolumn{1}{c}{}  & \multicolumn{1}{c}{} &  \multicolumn{1}{c}{} &  \multicolumn{1}{c}{}\\
\multicolumn{1}{l}{(1)} & \multicolumn{1}{c}{(2)} & \multicolumn{1}{c}{(3)} & \multicolumn{1}{c}{(4)} &  \multicolumn{1}{c}{(5)} & \multicolumn{1}{c}{(6)} & \multicolumn{1}{c}{(7)} & \multicolumn{1}{c}{(8)} & \multicolumn{1}{c}{(9)} & \multicolumn{1}{c}{(10)} & \multicolumn{1}{c}{(11)}  & \multicolumn{1}{c}{(12)} & \multicolumn{1}{c}{(13)} \\
\hline

 351 &  ATCA-ADFS J044823-532144 & 4:48:23.7 & -53:21:44.9 & 0.00 & 0.00 &   0.962 &   0.029 &   1.088 &   0.032 &  &  & \\
 352 &  ATCA-ADFS J044823-534125 & 4:48:23.8 & -53:41:25.6 & 0.04 & 0.04 &   0.414 &   0.061 &   0.670 &   0.069 &  &    2.8 &  38.4\\
 353 &  ATCA-ADFS J044824-531440 & 4:48:24.0 & -53:14:40.0 & 0.00 & 0.00 &   0.860 &   0.030 &   0.946 &   0.033 &  &  & \\
\hline
 \end{tabular}
\end{scriptsize}
\label{sourcecatalogueshort_8}
\end{table*}

\begin{table*}
\vspace{0pt}
\caption{Continuation of the source catalogue}
\begin{scriptsize}
\fontsize{8}{10}\selectfont
\begin{tabular}{l l r r r r r r r r r r r}
\hline
\multicolumn{1}{l}{No} & \multicolumn{1}{c}{Source name} & \multicolumn{1}{c}{RA} & \multicolumn{1}{c}{DEC} & \multicolumn{1}{c}{$\Delta$RA} & \multicolumn{1}{c}{$\Delta$DEC} & \multicolumn{1}{c}{S$_{\rm peak}$} & \multicolumn{1}{c}{$\Delta$S$_{\rm peak}$} & \multicolumn{1}{c}{S$_{\rm total}$} & \multicolumn{1}{c}{$\Delta$S$_{\rm total}$} & \multicolumn{1}{c}{$\theta_{maj}$} & \multicolumn{1}{c}{$\theta_{min}$} & \multicolumn{1}{c}{$PA$}\\ 
 \multicolumn{1}{l}{} &         & \multicolumn{1}{c}{h:m:s.s} & \multicolumn{1}{c}{d:m:s.s} & \multicolumn{1}{c}{${\prime\prime}$} & \multicolumn{1}{c}{${\prime\prime}$} & \multicolumn{1}{c}{mJy} & \multicolumn{1}{c}{mJy} & \multicolumn{1}{c}{mJy}  & \multicolumn{1}{c}{mJy}  & \multicolumn{1}{c}{$^{\prime\prime}$} &  \multicolumn{1}{c}{$^{\prime\prime}$} &  \multicolumn{1}{c}{$\ensuremath{^\circ}\,$}\\
  \multicolumn{1}{l}{} &         & \multicolumn{1}{c}{} & \multicolumn{1}{c}{} & \multicolumn{1}{c}{} & \multicolumn{1}{c}{} & \multicolumn{1}{c}{beam$^{-1}$} & \multicolumn{1}{c}{beam$^{-1}$} & \multicolumn{1}{c}{}  & \multicolumn{1}{c}{}  & \multicolumn{1}{c}{} &  \multicolumn{1}{c}{} &  \multicolumn{1}{c}{}\\
\multicolumn{1}{l}{(1)} & \multicolumn{1}{c}{(2)} & \multicolumn{1}{c}{(3)} & \multicolumn{1}{c}{(4)} &  \multicolumn{1}{c}{(5)} & \multicolumn{1}{c}{(6)} & \multicolumn{1}{c}{(7)} & \multicolumn{1}{c}{(8)} & \multicolumn{1}{c}{(9)} & \multicolumn{1}{c}{(10)} & \multicolumn{1}{c}{(11)}  & \multicolumn{1}{c}{(12)} & \multicolumn{1}{c}{(13)} \\
\hline

 401 &  ATCA-ADFS J044914-530716 & 4:49:14.7 & -53:07:16.9 & 0.00 & 0.00 &   0.560 &   0.046 &   0.710 &   0.047 &  &  & \\
 402 &  ATCA-ADFS J044917-532530 & 4:49:17.4 & -53:25:30.5 & 0.00 & 0.00 &   0.215 &   0.034 &   0.247 &   0.034 &  &  & \\
 403 &  ATCA-ADFS J044920-531748 & 4:49:20.2 & -53:17:48.7 & 0.04 & 0.06 &   0.123 &   0.029 &   0.263 &   0.031 &    7.7 &    3.4 & -18.5\\
\hline
 \end{tabular}
\end{scriptsize}
\label{sourcecatalogueshort_9}
\end{table*}

\begin{table*}
\vspace{0pt}
\caption{Continuation of the source catalogue}
\begin{scriptsize}
\fontsize{8}{10}\selectfont
\begin{tabular}{l l r r r r r r r r r r r}
\hline
\multicolumn{1}{l}{No} & \multicolumn{1}{c}{Source name} & \multicolumn{1}{c}{RA} & \multicolumn{1}{c}{DEC} & \multicolumn{1}{c}{$\Delta$RA} & \multicolumn{1}{c}{$\Delta$DEC} & \multicolumn{1}{c}{S$_{\rm peak}$} & \multicolumn{1}{c}{$\Delta$S$_{\rm peak}$} & \multicolumn{1}{c}{S$_{\rm total}$} & \multicolumn{1}{c}{$\Delta$S$_{\rm total}$} & \multicolumn{1}{c}{$\theta_{maj}$} & \multicolumn{1}{c}{$\theta_{min}$} & \multicolumn{1}{c}{$PA$}\\ 
 \multicolumn{1}{l}{} &         & \multicolumn{1}{c}{h:m:s.s} & \multicolumn{1}{c}{d:m:s.s} & \multicolumn{1}{c}{${\prime\prime}$} & \multicolumn{1}{c}{${\prime\prime}$} & \multicolumn{1}{c}{mJy} & \multicolumn{1}{c}{mJy} & \multicolumn{1}{c}{mJy}  & \multicolumn{1}{c}{mJy}  & \multicolumn{1}{c}{$^{\prime\prime}$} &  \multicolumn{1}{c}{$^{\prime\prime}$} &  \multicolumn{1}{c}{$\ensuremath{^\circ}\,$}\\
  \multicolumn{1}{l}{} &         & \multicolumn{1}{c}{} & \multicolumn{1}{c}{} & \multicolumn{1}{c}{} & \multicolumn{1}{c}{} & \multicolumn{1}{c}{beam$^{-1}$} & \multicolumn{1}{c}{beam$^{-1}$} & \multicolumn{1}{c}{}  & \multicolumn{1}{c}{}  & \multicolumn{1}{c}{} &  \multicolumn{1}{c}{} &  \multicolumn{1}{c}{}\\
\multicolumn{1}{l}{(1)} & \multicolumn{1}{c}{(2)} & \multicolumn{1}{c}{(3)} & \multicolumn{1}{c}{(4)} &  \multicolumn{1}{c}{(5)} & \multicolumn{1}{c}{(6)} & \multicolumn{1}{c}{(7)} & \multicolumn{1}{c}{(8)} & \multicolumn{1}{c}{(9)} & \multicolumn{1}{c}{(10)} & \multicolumn{1}{c}{(11)}  & \multicolumn{1}{c}{(12)} & \multicolumn{1}{c}{(13)} \\
\hline

 451 &  ATCA-ADFS J045052-531120 & 4:50:52.7 & -53:11:20.5 & 0.01 & 0.01 &   0.566 &   0.086 &   0.820 &   0.086 &  &  & \\
 452 &  ATCA-ADFS J045055-531247 & 4:50:55.6 & -53:12:47.5 & 0.00 & 0.00 &   0.823 &   0.104 &   0.578 &   0.104 &  &  & \\
 453 &  ATCA-ADFS J045102-524726 & 4:51:02.5 & -52:47:26.6 & 0.00 & 0.00 &   3.847 &   0.418 &   2.931 &   0.420 &  &  & \\
\hline
 \end{tabular}
\end{scriptsize}
\label{sourcecatalogueshort_10}
\end{table*}

\begin{table*}
\vspace{0pt}
\caption{Continuation of the source catalogue}
\begin{scriptsize}
\fontsize{8}{10}\selectfont
\begin{tabular}{l l r r r r r r r r r r r}
\hline
\multicolumn{1}{l}{No} & \multicolumn{1}{c}{Source name} & \multicolumn{1}{c}{RA} & \multicolumn{1}{c}{DEC} & \multicolumn{1}{c}{$\Delta$RA} & \multicolumn{1}{c}{$\Delta$DEC} & \multicolumn{1}{c}{S$_{\rm peak}$} & \multicolumn{1}{c}{$\Delta$S$_{\rm peak}$} & \multicolumn{1}{c}{S$_{\rm total}$} & \multicolumn{1}{c}{$\Delta$S$_{\rm total}$} & \multicolumn{1}{c}{$\theta_{maj}$} & \multicolumn{1}{c}{$\theta_{min}$} & \multicolumn{1}{c}{$PA$}\\ 
 \multicolumn{1}{l}{} &         & \multicolumn{1}{c}{h:m:s.s} & \multicolumn{1}{c}{d:m:s.s} & \multicolumn{1}{c}{${\prime\prime}$} & \multicolumn{1}{c}{${\prime\prime}$} & \multicolumn{1}{c}{mJy} & \multicolumn{1}{c}{mJy} & \multicolumn{1}{c}{mJy}  & \multicolumn{1}{c}{mJy}  & \multicolumn{1}{c}{$^{\prime\prime}$} &  \multicolumn{1}{c}{$^{\prime\prime}$} &  \multicolumn{1}{c}{$\ensuremath{^\circ}\,$}\\
  \multicolumn{1}{l}{} &         & \multicolumn{1}{c}{} & \multicolumn{1}{c}{} & \multicolumn{1}{c}{} & \multicolumn{1}{c}{} & \multicolumn{1}{c}{beam$^{-1}$} & \multicolumn{1}{c}{beam$^{-1}$} & \multicolumn{1}{c}{}  & \multicolumn{1}{c}{}  & \multicolumn{1}{c}{} &  \multicolumn{1}{c}{} &  \multicolumn{1}{c}{}\\
\multicolumn{1}{l}{(1)} & \multicolumn{1}{c}{(2)} & \multicolumn{1}{c}{(3)} & \multicolumn{1}{c}{(4)} &  \multicolumn{1}{c}{(5)} & \multicolumn{1}{c}{(6)} & \multicolumn{1}{c}{(7)} & \multicolumn{1}{c}{(8)} & \multicolumn{1}{c}{(9)} & \multicolumn{1}{c}{(10)} & \multicolumn{1}{c}{(11)}  & \multicolumn{1}{c}{(12)} & \multicolumn{1}{c}{(13)} \\
\hline

 501 &  ATCA-ADFS J045202-524914 & 4:52:02.2 & -52:49:14.8 & 0.02 & 0.03 &   7.654 &   1.169 &  18.004 &   1.257 &    7.2 &    5.4 &   3.1\\
 502 &  ATCA-ADFS J045203-525035 & 4:52:03.1 & -52:50:35.2 & 0.00 & 0.00 &  15.911 &   1.169 &   9.492 &   1.184 &  &  & \\
 503 &  ATCA-ADFS J045203-525347 & 4:52:03.2 & -52:53:47.6 & 0.04 & 0.10 &   3.673 &   0.485 &   5.336 &   0.556 &    7.8 &  &  22.8\\
\hline
 \end{tabular}
\end{scriptsize}
\label{sourcecatalogueshort_11}
\end{table*}

\end{document}